%% file: tos.tex
\documentclass[pre,twocolumn,showpacs,showkeys,superscriptaddress,preprintnumbers,floatfix]{revtex4-1}

\usepackage{etex}
\usepackage{ifpdf}
\usepackage{hyperref}
\usepackage{dcolumn}
\usepackage{url}
\usepackage{amsmath}
\usepackage{amscd}
\usepackage{amsfonts}
\usepackage{amssymb}
\usepackage{bm}   
\usepackage{bbm}
\usepackage{verbatim}
\usepackage{stmaryrd}
\usepackage{amsthm}
\usepackage{xcolor}
\usepackage{setspace}
\usepackage{braket}

\usepackage{tikz}
\usetikzlibrary{arrows}
\usetikzlibrary{automata}
\usetikzlibrary{decorations.pathreplacing}
\usetikzlibrary{positioning}
\usetikzlibrary{plotmarks}
\usetikzlibrary{calc}
\usetikzlibrary{patterns}
\usetikzlibrary{external}
\usepackage{pgfplots}
\pgfplotsset{compat=newest}

\usepackage{graphicx}
\usepackage{subfigure}

\theoremstyle{plain}    
\theoremstyle{plain}    
\theoremstyle{plain}    
\theoremstyle{plain}    
\theoremstyle{plain}    
\theoremstyle{plain}    
\theoremstyle{plain}    
\theoremstyle{plain}    
\theoremstyle{plain}    
\theoremstyle{plain}    
\theoremstyle{plain}    
\theoremstyle{plain}

\input{cmechabbrev}

\renewcommand{\H}{\operatorname{H}}

\newcommand{\Abet}{\mathcal{Y}}

\newcommand{\kB}{k_\text{B}}  

\begin{document}

\title{Correlation-powered Information Engines and\\
the Thermodynamics of Self-Correction}

\author{Alexander B. Boyd}
\email{abboyd@ucdavis.edu}
\affiliation{Complexity Sciences Center and Physics Department,
University of California at Davis, One Shields Avenue, Davis, CA 95616}

\author{Dibyendu Mandal}
\email{dibyendu.mandal@berkeley.edu}
\affiliation{Department of Physics, University of California, Berkeley, CA
94720, U.S.A.}

\author{James P. Crutchfield}
\email{chaos@ucdavis.edu}
\affiliation{Complexity Sciences Center and Physics Department,
University of California at Davis, One Shields Avenue, Davis, CA 95616}

\date{\today}
\bibliographystyle{unsrt}

\begin{abstract}

Information engines can use structured environments as a resource to generate
work by randomizing ordered inputs and leveraging the increased Shannon entropy
to transfer energy from a thermal reservoir to a work reservoir. We give a
broadly applicable expression for the work production of an information engine,
generally modeled as a memoryful channel that communicates inputs to outputs as
it interacts with an evolving environment. The expression establishes that an
information engine must have more than one memory state in order to leverage
input environment correlations. To emphasize this functioning, we designed an
information engine powered solely by temporal correlations and not
by statistical biases, as employed by previous engines. Key to this is the
engine's ability to synchronize---the engine automatically returns to a desired
dynamical phase when thrown into an unwanted, dissipative phase by corruptions
in the input---that is, by unanticipated environmental fluctuations. This
self-correcting mechanism is robust up to a critical level of corruption,
beyond which the system fails to act as an engine. We give explicit analytical
expressions for both work and critical corruption level and summarize engine
performance via a thermodynamic-function phase diagram over engine control
parameters. The results reveal a new thermodynamic mechanism based on
nonergodicity that underlies error correction as it operates to support
resilient engineered and biological systems.
\end{abstract}


\preprint{Santa Fe Institute Working Paper 2016-06-013}
\preprint{arxiv.org:1606.08506 [cond-mat.stat-mech]}

\maketitle


\setstretch{1.1}
\section{Introduction}

Intriguing connections between statistical mechanics and information theory
have emerged repeatedly since the latter's introduction in the 1940s.
Thermodynamic entropy in the canonical ensemble is the Shannon information of
the Boltzmann probability distribution \cite{Jayn57a}. Entropy production along
a dynamical trajectory is given by the relative entropy~\cite{Kawa07, Bata15}, an
information-theoretic quantity, of the forward trajectories with respect to the
time-reversed trajectories \cite{Croo98a}. Perhaps the most dramatic
connection, though, appears in the phenomenon of Maxwell's demon, a thought
experiment introduced by James C. Maxwell \cite{Maxw88a}. This is a
hypothetical, intelligent creature that can reverse the spontaneous relaxation
of a thermodynamic system, as mandated by the Second Law of thermodynamics, by
gathering information about the system's microscopic fluctuations and
accordingly modifying its constraints, \emph{without expending any net work}. A
consistent physical explanation can be obtained only if we postulate, following
Szilard \cite{Szil29a}, a thermodynamic equivalent of information processing:
Writing information has thermodynamic benefits whereas erasing information has
a minimum thermodynamic cost, $\kB T \ln2$ for the erasure of one bit of
information. This latter is Landauer's celebrated principle
\cite{Land61a,Benn82}.

The thermodynamic equivalent of information processing has the surprising
implication that we can treat the carrying capacity of an information storage
device as a thermodynamic fuel. This observation has led to a rapidly growing
literature exploring the potential design principles of nanoscale, autonomous
machines that are fueled by information. References \cite{Mand2013,Mand012a},
for example, introduced a pair of stochastic models that can act as an engine without heat dissipation
and a refrigerator without work expenditure, respectively.  These strange thermal devices are achieved by writing information on a tape of
``bits''---that is, on a tape of two-state, classical systems. A more realistic
model was suggested in Ref. \cite{Lu14a}. These designs have been extended to
enzymatic dynamics~\cite{Cao15}, stochastic feedback control~\cite{Shir15}, and
quantum information processing~\cite{Dian13, Chap15a}.

The information tape in the above designs can be visualized as a sequence of
symbols where each symbol is chosen from a fixed alphabet, as shown in
Fig.~\ref{fig:FullRatchet} for binary tape symbols. There is less raw information
in the tape if the symbols in the sequence are statistically correlated with
each other. For example, the sequence $\ldots101010\ldots$, consisting of
alternating $0$s and $1$s, encodes only a single bit of information on the
whole since there are only two such sequences (differing by a phase shift). Whereas, a sequence of $N$ random binary symbols encodes $N$ bits of
information. The thermodynamic equivalent of information processing,
therefore, says that we can treat the former (ordered) sequence as a
thermodynamic fuel. This holds even though it contains equal numbers of $0$s
and $1$s on average as in the fully random sequence, which provides no such fuel.

The design principles of \emph{information engines} \cite{Boyd14b} explored so
far, however, are not generally geared towards temporally correlated information tapes
~\cite{Mand012a, Mand2013, Stra2013, Bara2013, Hopp2014, Lu14a, Um2015,
Merh15a} since, by and large, only a tape's single-letter frequencies have been
considered. However, the existence of statistical correlations among the
symbols---that is, between environmental stimuli---is the rule, not an
exception in Nature. Even technologically, producing a completely
correlation-free (random) sequence of letters is a significant challenge
\cite{Park88, Jame90, Ferr92}. The thermodynamic value of statistical
correlations~\cite{Espo11, Saga2012b} and quantum entanglement~\cite{Oppe02,
Zure03, Maru05, Dill09, Dahl11, Jevt12, Funo13, Brag14, Pera15} have been
discussed widely in the literature. 
Our goal here is to extend the design of tape-driven
information engines to accommodate this more realistic scenario---information
engines that leverage temporally correlated environments to convert thermal energy to
useful work.

Other studies have taken a somewhat different approach to the description and
utilization of the thermodynamic equivalent of information processing.
References~\cite{Touc2000, Cao2004, Saga2010, Toya10a, Ponm2010, Horo2010,
Horo2011, Gran2011, Abre2011, Vaik2011, Abre2012, Kund2012, Saga2012b,
Kish2012, Um2015, Boyd16b} explored active feedback control of a stochastic
system by external means, involving measurement and feedback or measurement,
control, and erasure. While Refs.~\cite{Ito2013, Hart2014, Horo2014, Horo2015}
explored a multipartite framework involving a set of interacting, stochastic
subsystems and Refs.~\cite{Espo2012, Stra2013} studied steady-state models of
Maxwell's demon involving multiple reservoirs. And, finally,
Refs.~\cite{Horo2013, Bara2014b, Horo2014b} indicated how several of these
approaches can be combined into single framework.

Here, we use computational mechanics \cite{Crut12a} for thermal information
ratchets \cite{Boyd15a} to derive a general expression for work production that
takes into account temporal correlations in the environment as well as
correlations created in the output by the information engine's operation. The
functional form of the work expression establishes that memoryless information
ratchets cannot leverage anything more than single-symbol frequencies in their
input and are, therefore, insensitive to temporal correlations. Thus, to the
extent that it is possible to leverage temporally correlated environments,
memoryful information engines are the only candidates. This indicates, without
proof, that the memory of an information engine must reflect the memory of its
environment to most efficiently leverage structure in its input.

Adding credence to this hypothesis, we introduce an \emph{ergodic} information engine that is driven \emph{solely} by temporal correlations in the input symbols to produce work. The states of the engine wind up reflecting the memory states of the generator of the input process. This makes good on the conjecture \cite{Boyd15a} as to why one observes thermodynamically functional ratchets in the real world that support memory \cite{Boyd15a}: Only Demons with memory can leverage temporally correlated fluctuations in their environment.

Similar behavior was demonstrated by Maxwell's refrigerator
\cite{Mand2013}, when Ref. \cite{Chap15a} showed it to be a
nonergodic refrigerator when driven by a nonergodic process that is
statistically unbiased over all realizations.  However, we focus on our ergodic
engine, since ergodicity leads to robust and reliable work production.
This contrast is notable. Without ergodicity, an engine does not function during many realizations, from trial to trial. In this sense, a ``nonergodic engine'' is unreliable in performing its intended task, such as being an engine (converting thermal energy to work), generating locomotion, and the like.
During one trial it functions; on another it does not.

If one is willing to broaden what one means by ``engine", then one can imagine
constructing an ``ensemble engine'' composed of a large collection of
nonergodic engines and then only reporting ensemble-averaged performance.
Observed over many trials, the large trial-by-trial variations in work
production are masked and so the ensemble-average work production seems a
fair measure of its functionality. However, as noted, this is far from the
conventional notion of an engine but, perhaps, in a biological setting with
many molecular ``motors'' it may be usefully considered functional.

Our design of an ergodic engine that can operate solely on temporal
correlations should also be contrasted with a recent proposal~\cite{McGr16}
that utilizes mutual information between two tapes, i.e., \emph{spatial}
correlations, as a thermodynamic fuel.

The overarching thermodynamic constraints on functioning at all are analyzed in
a companion work \cite{Boyd16b}. The following, in contrast, focuses on the
particular functionality of self-correcting Demons in the presence of
temporally correlated environments and on analyzing the thermodynamic regimes
that support them. First, we review the information engine used and give a
synopsis of our main results so that they are not lost in the more detailed
development.  Second, the technical development begins as we introduce the
necessary tools from computational mechanics and stochastic thermodynamics.
Third, using them, we analyze the engine's behavior and functioning in the
presence of a correlated input, calling out the how the Demon recognizes (or
not) correlations in the input and either (i) responds constructively by using
them to convert thermal energy to work or (ii) dissipates energy as it attempts
to re-synchronize and regain engine functioning. Fourth, we note how these two
dynamical modes represent a type of dynamical nonergodicity over the ratchet's
state space when the ratchet cannot re-synchronize, which leads to temporary
nonergodicity in the work production. However, with re-synchronization, these
two dynamical modes become accessible from each other, which leads to
ergodicity of the engine and its work production. And, finally, we derive the
physical consequences for the costs of self-correction and its operational
limits.

\section{A Self-Correcting Information Engine: Synopsis}

\begin{figure}[tbp]
\centering
\includegraphics[width=\columnwidth]{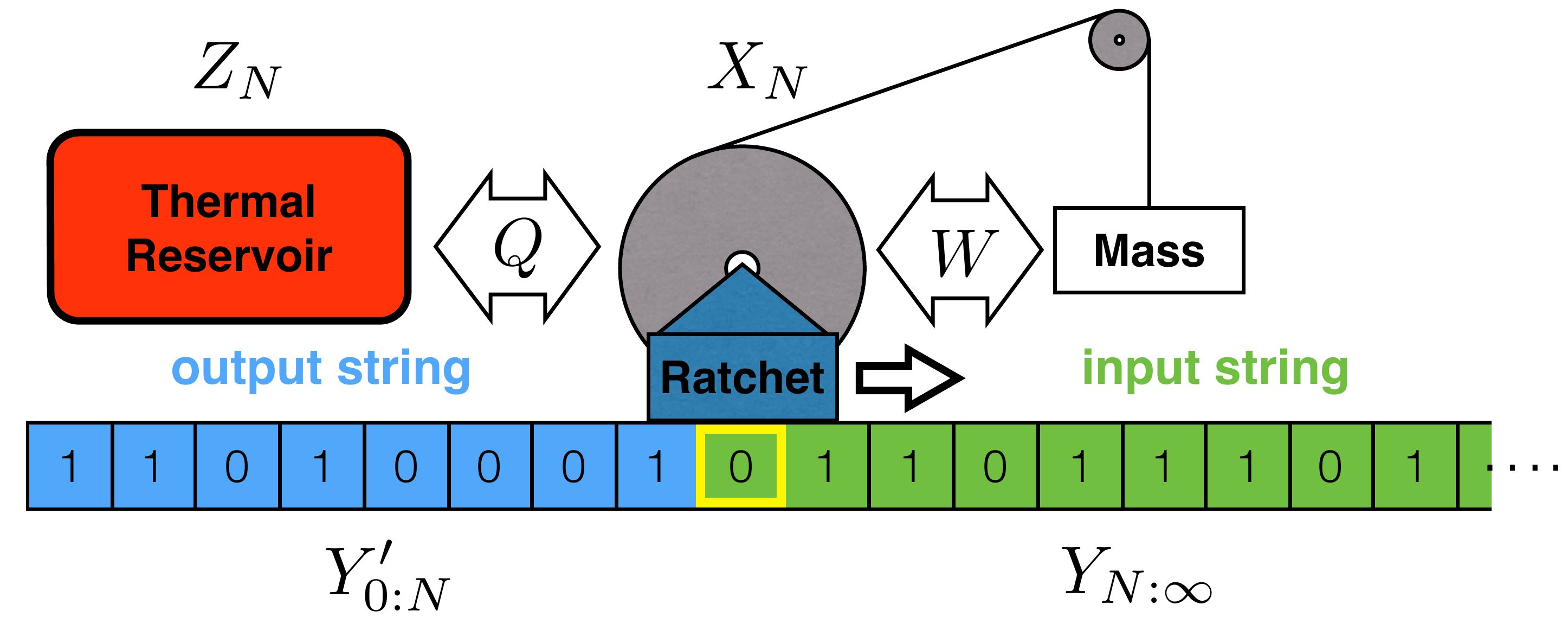}
\caption{\textbf{Thermal ratchet information engine:} A ratchet and three
	reservoirs---work, heat, and information. The work reservoir is depicted as
	gravitational mass suspended by a pulley. The information reservoir
	consists of a string of cells, each a two-state classical system that
	encodes one bit of information. The ratchet moves unidirectionally along
	the string and exchanges energy between the heat and the work reservoirs.
	The ratchet reads the value of a single cell (highlighted in yellow) at a
	time from the input string (green, right), interacts with it, and writes a
	symbol to the cell in the output string (blue, left) of the information
	reservoir. Information exchange between the ratchet and the information
	reservoir is signified by the change in the information content of the
	output symbols with respect to the input symbols. Driven by the information
	exchange, the ratchet transduces the input string $Y_{0:\infty}=Y_0Y_1...$
	into an output string $Y'_{0:\infty}=Y'_0Y'_1 \ldots$.
	(Reprinted from Ref. \protect\cite{Boyd15a} with permission.)
	}
\label{fig:FullRatchet}
\end{figure}

Figure~\ref{fig:FullRatchet} shows our model \cite{Mand012a,Boyd15a} of an
information engine implemented as a thermal ratchet consisting of four
elements: a thermal reservoir, a work reservoir (mass in a gravitational field),
an information tape (or reservoir), and a ratchet controlled by the values in
the input tape cells. The ratchet acts as the communication medium between the
three reservoirs as it moves along the tape and transforms the input
information content. In the process, it mediates energy exchange between the
heat and work reservoirs.

To precisely specify the kinds of temporal correlation in the ratchet's
environment, we represent the generator of the sequences on the information
tape via a hidden Markov model (HMM), a technique introduced in
Ref.~\cite{Boyd15a}. This has several advantages. One is that the full
distribution over infinite sequences of the input tape
$\Pr(\overleftrightarrow{Y})$ is represented in a compact way. The most extreme
case of this comes in recalling that finite-state HMMs can finitely represent
infinite-order Markov processes \cite{Jame10a}. And so, HMMs give a desirable
flexibility in the kinds of environments we can analyze, from memoryless to
finite- and infinite-order Markovian. Another is that many statistical and
informational properties can be directly calculated, as we discuss shortly. In
this setup, the ratchet is a transducer in the sense of computational mechanics
~\cite{Barn13a}. And this, in turn, allows exact analysis of informational
bounds on work production~\cite{Boyd15a,Merh15a}. Here, though, in Sec.
\ref{sec:TransducerEntropy} we go further, expanding the toolset of the
HMM-transducer formalism by deriving a general work production expression for
any finite-state input HMM driving a finite-state thermal ratchet.

With this powerful new expression for work production, Sec. \ref{sec:P2Input}
then considers the case of a perfectly correlated information tape. Though
nominally simple, this case is of particular interest since previous
single-symbol entropy bounds erroneously suggest this class of input should
generate nonpositive work. Our entropy rate bounds, in contrast, suggest it is
possible to generate net positive work. And, indeed, we see that the
single-symbol bounds are violated, as our ratchet produces positive work. In
examining this concrete model, moreover, we realize that the ratchet's
synchronizing to the correlations in its input is an essential part of work
production: Synchronization is how the ratchet comes to leverage the
thermodynamic ``fuel'' in a memoryful input process.

This result emphasizes a key feature of our ratchet design: Useful
thermodynamic functioning is driven purely by the temporal correlations in the
input tape. That is, if the symbols are perfectly correlated---a sequence with
temporal memory, e.g., with $1$s always following $0$s and vice versa---the
ratchet acts as an engine, writing new information on the output tape and
transferring energy from the heat to the work reservoir. However, if the
correlation is not perfect, depending on engine parameters, the ratchet can act
as an information-eraser or dud, converting work into heat. Thus, there exists
a critical level of corrupted input correlation beyond which engine
functionality is no longer possible. Our tools allow us to give explicit
expressions for work in all these cases, including the parameter limits of
thermodynamic functioning.

Perhaps most importantly, the analysis reveals a novel mechanism underlying the
functioning and its disappearance. This can be explained along the following
lines. An exclusive feature of the ratchet design is the presence of a
synchronizing state, denoted $C$ in the (state $\otimes$ bit)-transition
diagram of Fig.~\ref{fig:DetailedBalancedMarkov}. Absent $C$ and for perfectly
correlated input, the ratchet is equally likely to be in two stable dynamical
modes: ``clockwise'' in which heat is converted into work and
``counterclockwise'' in which work is converted into heat. (See
Fig.~\ref{fig:PathDiagram}.) Since the counterclockwise mode dissipates more
per cycle than can be compensated by the clockwise mode, without $C$ the
ratchet cannot function as an engine. With $C$, though, the counterclockwise
mode becomes a transient and the clockwise mode an attractor, making possible
the net conversion of heat into work (engine mode). The phenomenon of an
observer (ratchet) coming to know the state of its environment (phase of the
memoryful input tape) is referred to as \emph{synchronization} \cite{Crut01a}.
(For a rather different notion of synchronization and its thermodynamic
interpretation see Ref.~\cite{Izum16}.)

In contrast, when the input symbols are not perfectly correlated due to
phase slips, say, the ratchet is randomly thrown into the dissipative
counterclockwise mode. Nonetheless, repeated re-synchronization may compensate,
allowing the engine mode, if the transition probabilities into $C$ are
enhanced, up to a level. This is a form of \emph{dynamical error correction}.
Beyond a certain level of corruption in the input correlations, however,
dynamical error correction is not adequate to resynchronize to the input phase.
The Demon cannot act as an engine, no matter how large the transition
probabilities into $C$. This critical corruption level is shown in the
thermodynamic-function diagram of Fig.~\ref{fig:PhaseDiagram} by the
vertical dotted line, where the horizontal axis denotes level of corruption
as the frequency of phase slips.

The current situation must be contrasted with the usual error correction
schemes in communication theory and biological copying. In the former context,
redundancy is built into the data to be transmitted so that errors introduced
during transmission can be corrected by comparing to redundant copies, up to a
certain capacity. In the biological context of copying,  as in DNA
replication~\cite{Andr08}, error correction corresponds to the phenomenon of
active reduction of errors by thermodynamic means~\cite{Hopf74, Nini75,
Ehre80}. In the current context, we use the term {\it self-correction} to refer
to the fact the proposed information engine can predict and synchronize itself
with the state of the information source to produce positive work even when the
engine is initiated in or driven by fluctuations to a dissipative mode.
Section~\ref{sec:CorruptInput} discusses this self-correcting behavior of the
engine in detail.

To analyze how dynamical error correction operates quantitatively, the
following shows how the presence of state $C$ renders the counterclockwise
phase transient. This reveals a novel three-way tradeoff between
synchronization rate (transition probability from $C$ to the clockwise phase),
work produced during synchronization, and average extracted work per cycle.
Section \ref{sec:CorruptInput} then turns to analyze re-synchronization,
considering the case of imperfectly correlated information tape with phase
slips. It demonstrates how the ratchet dynamically corrects itself and converts
heat into work over certain parameter ranges. The section closes by giving the
expression for maximum work and the parameter combinations corresponding to
achieving optimum conversion.

Throughout the exploration, several lessons stand out. First, to effectively
predict bounds on a input-driven ratchet's work production, one must consider
\emph{Shannon entropy rates} of the input and output strings; and not
single-variable entropies. Second, the expression for the work production shows
that correlations coming from memoryful environments can only be leveraged by
memoryful thermodynamic transformations (Demons). While it remains an open
question how to design ratchets to best leverage memoryful inputs, the
particular ratchet presented here demonstrates how important it is for the
ratchet's structure to ``match'' that of the input correlations. In short, the
ratchet only produces work when its internal states are synchronized to the
internal states of the input sequence generator. Otherwise, it is highly
dissipative. And last, synchronization has energetic consequences that
determine the effectiveness of dynamical error correction and the tradeoffs
between average work production, work to synchronize, and synchronization rate.

\section{Thermal Ratchet Principles}
\label{sec:Setup}

Our finite-state ratchet, shown above in Fig. \ref{fig:FullRatchet}, moves
along the information tape unidirectionally, interacting with each symbol
sequentially. The ratchet interacts with each symbol for time $\tau$ and
possibly switches the symbol value contained in the cell. We refer to time
period $\tau$ as the \emph{interaction interval} and the transitions that
happen in the joint state space as \emph{interaction transitions}. Through this
process, the ratchet transduces a semi-infinite input string, expressed by
random variable $Y_{0:\infty} = Y_0 Y_1 \ldots$, into an output string
$Y'_{0:\infty} = Y'_0 Y'_1 \ldots$. Here, the symbols $Y_N$ and $Y'_N$ realize
the elements $y_N$ and $y'_N$, respectively, over the same information alphabet
$\mathcal{Y}$.

For example, as in Fig.~\ref{fig:FullRatchet}, the alphabet consists of just
$0$ and $1$. Consider the case in which the ratchet was initiated at the
leftmost end at time $t = 0$. At time $t = N \tau$ the entire tape is described
by the random variables $Y'_{0:N} Y_{N\infty} = Y'_0 Y'_2 \ldots Y'_{N-2}
Y'_{N-1} Y_N Y_{N+1} \ldots$, because in $N$ time-steps $N$ input symbols have
been transduced into $N$ output symbols. The state of the ratchet at time $t =
N \tau$ is denoted by the random variable $X_N$, which realizes an element $x_N
\in \mathcal{X}$, where $\mathcal{X}$ is the ratchet's state space.

Since we chose the input alphabet to consist of just two symbols $0$ and $1$, we
refer to the values in the tape cells as \emph{bits}. That this differs from the
information unit ``bit'' should be clear from context. \emph{Tape} generally
refers to the linear chain of cells and \emph{string} to the stored sequence of symbols or cell values.

Finally, the ratchet is connected with two, more familiar reservoirs---a
thermal reservoir and a work reservoir. The state of the thermal reservoir at
time $t = N\tau$ is denoted by $Z_N$. We assume that the thermal reservoir is
at absolute temperature $T$ K. The work reservoir consists of a mass being
pulled down by gravity, but kept suspended by a pulley. Certain, specified
ratchet transitions lower and raise the mass, exchanging work.

To set up the analysis, we must first review how to measure information,
structure, and energy as they arise during the ratchet's operation.

\begin{figure}[tbp]
\centering
\includegraphics[width=\columnwidth]{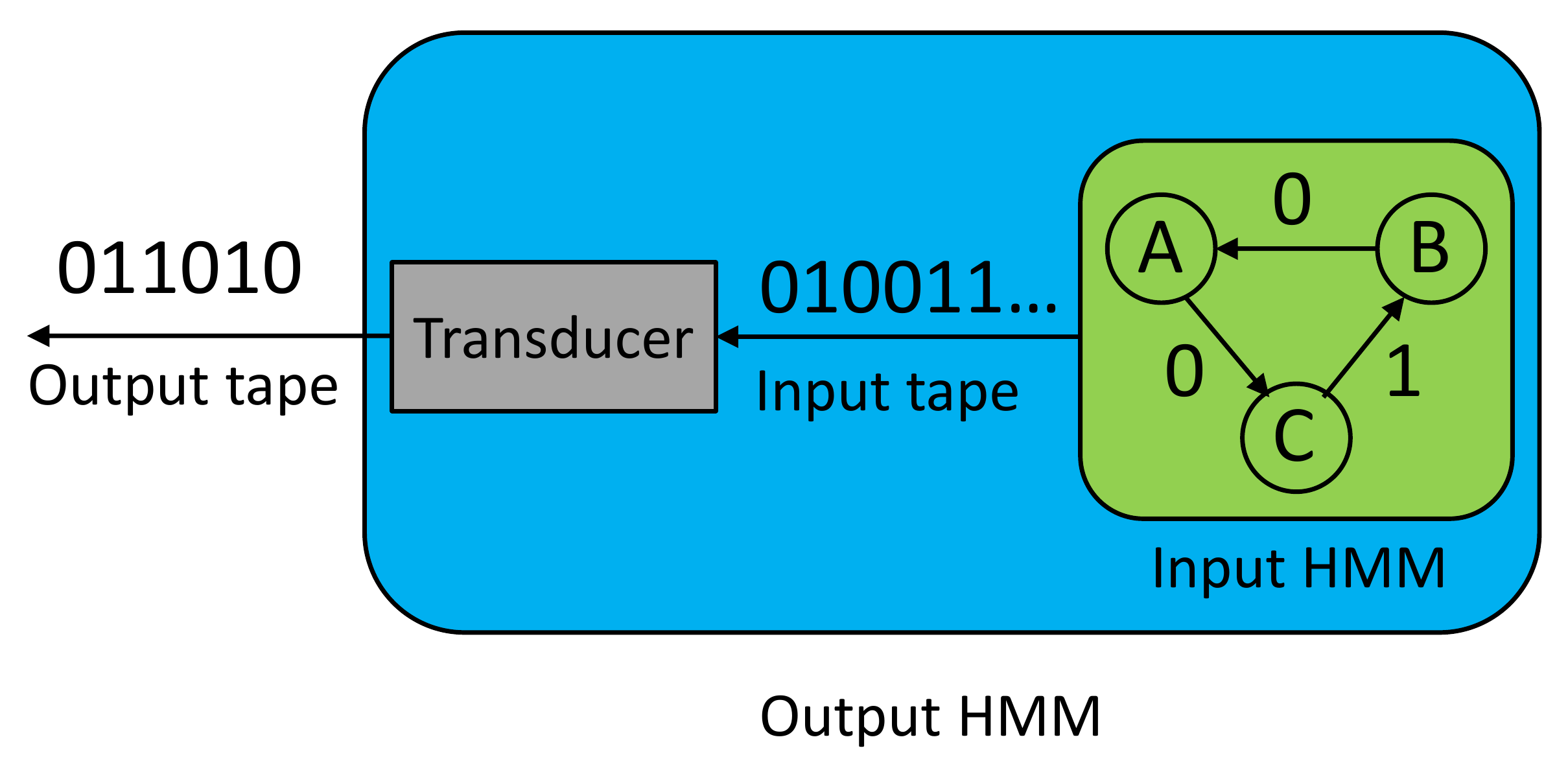}
\caption{{\bf Computational mechanics of information engines:} The input tape
	values are generated by a hidden Markov model (HMM) with, say, three hidden
	states---$A$, $B$, and $C$. Specifically, transitions among the hidden
	states produce $0$s and $1$s that form the input tape random variables
	$Y_{0:\infty}$. The ratchet acts as an informational transducer that
	converts the input HMM into an output process, that is also represented as
	an HMM. That is, the output tape $Y'_{0:\infty}$ can be considered as
	having been generated by an effective HMM that is the composition of the
	input HMM and the ratchet's transducer \cite{Barn13a}.
	}
\label{fig:CompMechPic}
\end{figure}

\subsection{Ratchet Informatics: Computational Mechanics}
\label{sec:CompMech}

To monitor information generation and storage, computational mechanics
views the sequence of symbols from the left of the input tape $Y_{0:\infty}$
as the temporal output of a kind of HMM, called an \eM\ \cite{Crut12a}.
The latter provides the most compact way to represent the statistical
distribution of symbol sequences. In particular, many types of long-range
correlation among the symbols are encoded in the \eM's finite-state hidden
dynamics. The correlations appear as the \emph{memory}, characterized by its
internal-state entropy or \emph{statistical complexity} $\Cmu$. Specifically, if the input can be produced by an HMM with a single hidden state, the input generator is memoryless and there cannot be any correlation among the symbols
\footnote{Thus, in our use of the descriptor ``correlated'', the all $0$s
sequence and the all $1$s sequence have no temporal correlation. Since their
internal memory $\Cmu = 0$, they have no information to correlate. This is analogous to \emph{autocorrelation} in which the zero frequency offset is subtracted.}.

The ratchet functions as a memoryful communication channel that sequentially
converts the input symbols into values in $Y'_{0:\infty}$, the output tape.
Naturally, the output tape itself can be considered in terms of another HMM, as
emphasized by the schematic in Fig.~\ref{fig:CompMechPic}. There, the ratchet
acts as an information transducer between two information sources represented
by respective input and output HMMs~\cite{Barn13a}.

These choices make it rather straightforward to measure ratchet \emph{memory}.
If the size of its state space is unity ($|\mathcal{X}|=1$), then we say it is
memoryless. Otherwise ($|\mathcal{X}|>1$), we say it is memoryful. With memory,
the ratchet at time $t = N\tau $ can store information about the past input
symbols $y_{0:N}$ with which it has interacted, as well as past outputs
$y'_{0:N}$. Similarly, the output HMM can have memory (its own positive
statistical complexity $\Cmu > 0$) even when the input HMM does not. This was
the case, for example in Refs. \cite{Mand012a,Mand2013, Lu14a,Boyd15a}.
Critically, the transducer formalism has the benefit that we can exactly
calculate the distribution $\Pr(Y'_{0:\infty})$ of output tapes for any
finite-memory ratchet with a finite-memory input process. Shortly, we add to
this set of tools, introducing a method to calculate the work production by any
finite-memory ratchet operating on a finite-memory input.

\subsection{Ratchet Energetics: The First Law of Thermodynamics}
\label{sec:TransducerEnergy}

Interactions between ratchet states and input symbols have energetic consequences. The internal states and symbols interact with a thermal reservoir at temperature $T$, whose configuration at time step $N$ is denoted by the random variable $Z_N$, and with a work reservoir, that holds no information and so need not have an associated random variable. Through its operation, the current input symbol facilitates or inhibits energy flows between the work and thermal reservoirs.

The joint dynamics of the ratchet and incoming symbol occur over two
alternating steps: a \emph{switching} transition and an \emph{interaction}
transition. At time $t = N \tau$, the ratchet switches the tape cell with which
it interacts from the $(N-1)$th output symbol $y'_{N-1}$ to the $N$th input
symbol $y_{N}$. This is followed by the interaction transition between the
ratchet, which is in the $x_N$ state, and the symbol $y_{N}$. Together, they
make a stochastic transition in their joint state space according to the Markov
chain:
\begin{align*}
& M_{x_N\otimes y_N \rightarrow x_{N+1} \otimes y'_N}= \\
  & \quad \Pr(X_{N+1}=x_{N+1},Y'_{N}=y'_N|X_N=x_N,Y_N=y_N)
  ~.
\end{align*}
$M$ has detailed balance, since
transitions are activated by the thermal reservoir. Energy changes due to
these thermal interaction transitions are given by the Markov chain:
\begin{align*}
\Delta E_{x_N \otimes y_N \rightarrow x_{N+1} \otimes y'_N}
  = k_B T \ln
  \frac{M_{x_{N+1} \otimes y'_N \rightarrow x_{N} \otimes y_N}}
  {M_{x_N \otimes y_N \rightarrow x_{N+1} \otimes y'_N}}
  ~.
\end{align*}
These energies underlie the heat and work flows during the ratchet's
operation.  Through interaction, the input symbol $y_{N}$ is converted into the
output symbol $y'_{N}$ and written to the output tape cell as the ratchet
switches to the next input bit $y_{N+1}$ to start the next interaction at time
$t=(N+1)\tau$.

Notably, previous treatments \cite{Mand012a, Mand2013, Boyd15a} of information
engines associated the energy change during an interaction transition with work
production by coupling the interaction transitions to work reservoirs. While it
is possible to construct devices that have this work generation scheme, it
appears to be a difficult mechanism to implement in practice. We avoid this
difficulty, designing the energetics in a less autonomous way, not attaching
the work reservoir to the ratchet directly.

So, instead of the ratchet effortlessly stepping along the tape
unidirectionally on its own, it is driven. (And, an energetic cost can be
included for advancing the ratchet without loss of generality.) In this way,
heat flow happens during the interaction transitions and work flow happens
during the switching transitions.  Appendix
\ref{app:GeneralizedRatchetEnergetics} shows how this strategy gives an exact
asymptotic average work production per time step:
\begin{align}
\langle W \rangle =
  \sum_{\substack{x,x' \in \mathcal{X} \\ y, y' \in \mathcal{Y}}}
  \pi_{x \otimes y} M_{x \otimes y \rightarrow x' \otimes y'}
  \Delta E_{x \otimes y \rightarrow x' \otimes y'}
  ~,
\label{eq:Work}
\end{align}
where $\pi_{x\otimes y}$ is the asymptotic distribution over the joint state of
the Demon and interaction cell at the beginning of any interaction transition:
\begin{align}
\pi_{x \otimes y} = \lim_{N \rightarrow \infty} \Pr(X_N=x,Y_N=y)
  ~.
\label{eq:Pi}
\end{align}  
It is important to note that $\pi$ is not $M$'s stationary distribution and,
moreover, it is highly dependent on the input HMM. Despite calculating work
production for a different mechanism, the asymptotic power calculated here is
the same as in previous examinations ~\cite{Mand012a,Boyd15a,Merh15a}.

From the expression of work given in Eq.~(\ref{eq:Work}), we see that
memoryless ratchets have severe limitations in their ability to extract work
from the heat reservoir. In this case, the ratchet state space $\mathcal{X}$
consists of a single state and $\pi$ in Eq.~(\ref{eq:Pi}) is just the single
symbol distribution of the input string:
\begin{align*}
\pi_{x \otimes y}=\Pr(Y_0=y)
  ~.
\end{align*}
As a result, the calculation of work depends only on the single-symbol
statistics of the input string, producing work from the string as if the input
were independent and identically distributed (IID). Regardless of whether there
are correlations among the input symbols, the work production of a memoryless
ratchet is therefore the same for all inputs having the same single-symbol
statistics. For example, a memoryless ratchet cannot distinguish between input
strings $01010101\ldots$ and $00110011\ldots$ as far as work is concerned.
Thus, for the ratchet to use correlations in the input string to generate work,
it must have nonzero memory. This is in line with previous examinations of
autonomous information engines ~\cite{Merh15a,Boyd15a}. In any case, the
general form for the work production here allows one to calculate it for any
finite memoryful channel operating on any input tape generated by a finite HMM.

\subsection{Ratchet Entropy Production: The Second Law of Thermodynamics}
\label{sec:TransducerEntropy}

Paralleling Landauer's Principle \cite{Land61a,Benn82} on the thermodynamic
cost of information erasure, several extensions of the Second Law of
thermodynamics have been proposed for information processing. We refer to them
collectively as the \emph{thermodynamic equivalents of information processing}.
For ratchets, these bounds on the thermodynamic costs of information
transformation can be stated either in terms of the input and the output HMMs'
\emph{single-symbol entropy} (less generally applicable) or \emph{entropy rate}
(most broadly applicable). Let's review their definitions for the sake of
comparison.

Consider the probability distribution of the symbols $\{0, 1\}$ in the output
sequence of an HMM. If the single-symbol probabilities are $\{p, 1-p\}$,
respectively, the single-symbol entropy $\H_1$ of the HMM is given by the
\emph{binary entropy function} $\H(p)$ \cite{Cove06a}:
\begin{align}
\label{eq:SingleSymbol}
\H_1 & = \H (p) \\
     & \equiv - p \ln{p} - (1- p) \ln{(1- p)} \nonumber
  ~. 
\end{align}
By definition, single-symbol entropy ignores sequential symbol-symbol
correlations.

The entropy rate, in contrast, is the asymptotic per-symbol uncertainty. To
define it, we need to first introduce the concept of a word in the output
sequence generated by an HMM. A word $w$ is a subsequence of symbols of length
$\ell$ over the space $\Abet^\ell$. For example, a binary word of length $\ell
= 2$ consists of a pair of consecutive symbols; an event in the space $\Abet^2
= \{00, 01, 10, 11\}$. Thus, there are $2^\ell$ possible length-$\ell$ words or
elements in $\Abet^\ell$. The \emph{Shannon entropy rate} of the process
generated by an HMM is then given by \cite{Cove06a}:
\begin{align}
\hmu = - \lim_{\ell \to \infty} \frac{1}{\ell}
	\sum_{w \in \Abet^\ell} \Pr (w) \ln_2 {\Pr (w)}
	~,
\label{eq:EntropyRate}
\end{align} 
where $\Pr (w)$ denotes the probability of $w \in \Abet^\ell$. Entropy rate
$\hmu$ captures the effects of correlations in the symbols at all lengths.

For memoryless processes, $\H_1 = \hmu$. Otherwise, $\H_1 > \hmu$, with $\hmu$
being the correct measure of information per symbol and $\H_1$ being an
overestimate. One relevant extreme case arises with exactly periodic processes
with period greater than $1$: $\hmu = 0$; whereas $\H_1 > 0$, it's magnitude
being determined by the single-symbol frequencies.

We can now state two specific forms of the thermodynamic equivalent of
information processing for information engines:
\begin{align}
\label{eq:SecondLaw1}
\langle W \rangle & \leq k_B T \, \ln 2 \, \Delta \H_1 \\
\label{eq:SecondLaw2}
\langle W \rangle & \leq k_B T \, \ln 2 \, \Delta \hmu
  ~,
\end{align}
where $\Delta \H_1$ and $\Delta h_\mu$ denote, respectively, the change in
single-symbol entropy and in entropy rate from the input HMM to the output HMM~\cite{Mand012a,Mand2013,Bara2013, Deff2013, Bara2014b,Bara2014a,Merh15a,Boyd15a}. 

Let's compare them. Equation~(\ref{eq:SecondLaw1}) says that correlations in
the input string beyond single symbols cannot be used to produce work, while
Eq.  (\ref{eq:SecondLaw2}) suggests that it is possible. This follows since,
if we keep the single-symbol probabilities constant while increasing the
temporal correlations in the input, all while keeping the output fixed, $\Delta
\H_1$ remains constant, but $\Delta h_\mu$ increases.

To resolve this seeming ambiguity, we appeal to the general expression of
Eq.~(\ref{eq:Work}) for calculating work production. The expression says that
work production depends on the memory of both the ratchet and the input HMM;
see App. \ref{app:GeneralizedRatchetEnergetics}. In this way, temporal
correlations in the input string can influence the ratchet's thermodynamic
behavior. Only when the ratchet is memoryless is there no relevance of the
correlations, so far as the average work is concerned. In the memoryless case,
Eq.~(\ref{eq:SecondLaw1}) as well as Eq.~(\ref{eq:SecondLaw2}) are valid.

This observation suggests that, in contrast, for a ratchet to use correlations
in the input string to generate work, it must have more than one internal
state~\cite{Boyd16b}. In addition, to generate correlations in the input
string, its generating HMM must have memory.  This leads to the intuitive
hypothesis that to leverage work from the temporal order in the input string
(correlations created by the input HMM's memory), the ratchet must also have
memory.

We test this hypothesis by analyzing the specific example of a perfectly
correlated environment---a periodic input process. As we do, keep in mind that,
on the one hand, Eq. (\ref{eq:SecondLaw1}) says that no work production is
possible, regardless of the binary output process statistics. On the other
hand, Eq. (\ref{eq:SecondLaw2}) suggests the opposite. As long as the output
process has some uncertainty in sequential symbols, then $\Delta h_\mu > 0$. We also introduce a
ratchet with three memory states that produces positive work and even appears
to be nearly optimal for certain parameter ranges~\cite{Boyd16b}. In short, a
memoryful ratchet with a memoryful input process violates Eq.
(\ref{eq:SecondLaw1}), demonstrating that bound's limited range of application.

\section{Functional Ratchets in Perfectly Correlated Environments}
\label{sec:RatchetCorrelatedEnv}

Let's consider the case of a correlated environment and then design a thermal
ratchet adapted to it.

\subsection{The Period-$2$ Environment}
\label{sec:P2Input}

\begin{figure}[tbp]
\centering
\includegraphics[width=.5\columnwidth]{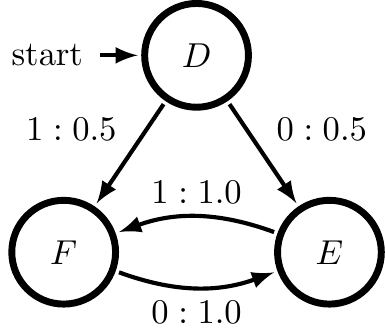}
\caption{{\bf Period-$2$ process hidden Markov model} with a transient start
	state $D$ and two recurrent causal states $E$ and $F$. Starting from $D$,
	the process makes a transition either to $E$ or to $F$ with equal
	probabilities while emitting $y = 0$ or $y = 1$, respectively. This is
	indicated by the transition labels from $D$: $y:p$ says generate symbol $y$
	when taking the transition with probability $p$. On arriving at states $E$
	or $F$, the process alternates between two states, emitting $y = 0$ for
	transitions $E \rightarrow F$ and $y = 1$ for transitions $F \rightarrow
	E$. In effect, we get either of two infinite sequences, $y_{0:\infty} =
	0101 \ldots$ and $y_{0:\infty} = 1010\ldots$, with equal probabilities.
	}
\label{fig:Period2}
\end{figure}

Take the specific case of a period-$2$ input process. The state
transition diagram for its HMM is given in Fig.~\ref{fig:Period2}. There are
three internal states. $D$ is a transient state from which the process starts.
From $D$, the process transitions to either $E$ or $F$ with equal
probabilities. If the system transitions to $E$, a $0$ is emitted, and if the
system transitions to $F$, a $1$ is. Afterwards, the process switches between
$E$ and $F$ with $E \rightarrow F$ transitions emitting $1$ and $F \rightarrow
E$ transitions emitting $0$. As a result, the input HMM generates two possible
sequences that drive the ratchet: $y_{0:\infty} =010101 \ldots$ or
$y_{0:\infty} = 101010 \ldots$. Note that these two sequences differ by a
single phase shift.

The period-$2$ process is an ideal base case for analyzing how ratchets extract
work out of temporal correlations. First, its sequences have no bias in the
frequencies of $0$'s and $1$'s, as they come in equal proportions; thereby
removing any potential gain from an initial statistical bias. And, second, the
symbols in the sequence are perfectly correlated---a $0$ is followed by $1$ and
a $1$ by $0$.

More to the point, previous information engines cannot extract work out of such
periodic sequences since those engines were designed to obtain their
thermodynamic advantage purely from statistical biases in the inputs
\cite{Mand012a, Mand2013, Bara2013, Boyd15a, Merh15a}. By way of contrast, we
now introduce and analyze the performance of a ratchet design that extracts
work out of such perfectly correlated, unbiased input sequences. The following
section then considers the more general case in which input correlations are
corrupted by environmental fluctuations.

Let's explain the information-theoretic reasoning that motivates this. For a
period-$2$ process, the single-symbol entropy $\H_1$ is maximal: $\H[Y_N] = 1$.
However, its entropy rate $\hmu = 0$ due to its perfect predictability as soon
as any symbol is known. This, on the one hand, implies $\Delta \H_1 \equiv
\H[Y'_N] - \H[Y_N] \leq 0$. Equation~(\ref{eq:SecondLaw1}), in turn, says that
work cannot be extracted regardless of the realizations of the output string;
no matter the design of the information engine. For the period-$2$ input,
though, $\Delta h_\mu = h'_\mu \geq 0$. And, Eq.~(\ref{eq:SecondLaw2})
indicates that work can be extracted as long as the output string has nonzero
entropy rate $h'_\mu$. This is achievable with appropriate thermal ratchet
design. In other words, Eq.~(\ref{eq:SecondLaw1}) suggests that it is
impossible to extract work from input correlations beyond single-symbol bias,
while Eq.~(\ref{eq:SecondLaw2}) suggests it is possible. We resolve this
disagreement in favor of Eq.~(\ref{eq:SecondLaw2}) by explicit construction
and exact analysis.

\begin{figure}[tbp]
\centering
\includegraphics[width=\columnwidth]{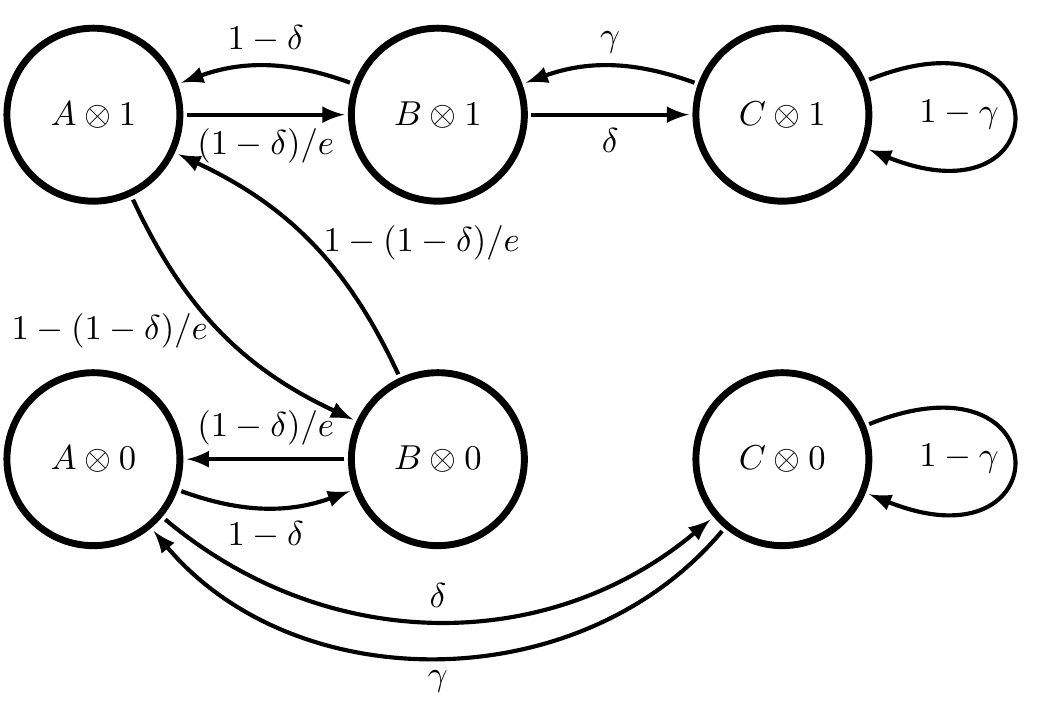}
\caption{\textbf{State transition diagram of a ratchet that extracts work out
	of environment correlations}: $A$, $B$, and $C$ denote the ratchet's
	internal states and $0$ and $1$ denote the values of the interacting cell.
	The joint dynamics of the Demon and interacting cell take place over the
	space of six internal joint states: $\{$A$\otimes$0, $\ldots$ ,
	C$\otimes$1$\}$. Arrows indicate the allowed transitions and their
	probabilities in terms of the ratchet control parameters $\delta$ and
	$\gamma$.
	}
\label{fig:DetailedBalancedMarkov}
\end{figure}

\subsection{Memoryful Ratchet Design}
\label{seubsec:MemoryfulRatchet}

Figure \ref{fig:DetailedBalancedMarkov} gives a ratchet design that can extract
work out of a period-$2$ process. As explained above in Sec.~\ref{sec:Setup},
the ratchet interacts with one incoming symbol at a time. As a result, the
ratchet's transducer specifies both ratchet internal states and the states of
the input tape cell being read. In the figure, $A$, $B$, and $C$ denote the
ratchet's internal states and $x \otimes y$ denotes the joint transducer state
of the ratchet state and interacting cell value, with the ratchet being in
state $x \in \{A, B, C\}$ and the interacting cell with value $y \in \{0,1\}$.
Arrows denote the allowed transitions and their labels the transition
probabilities in terms of ratchet control parameters, that we now introduce.
For example, if the Demon is in state $A$ and the input symbol has value $0$,
they make a transition to the joint state $B \otimes 0$ with probability $(1-
\delta)$ or to the joint state $C \otimes 0$ with probability $\delta$. Due to
conservation of probability, the sum of transition probabilities out of any
joint state is unity. After the transition, the old symbol value in the tape
cell is replaced by a new value. If the joint state made a transition to $B
\otimes 0$ and the incoming symbol had value $1$, the joint state is switched
to $B \otimes 1$. Then, a transition from joint state $B \otimes 1$ takes place
according to the rule described above.

The parameters $\delta$ and $\gamma$ satisfy the following constraint: $0 \leq
\delta, \gamma \leq 1$. The Markov chain matrix $M$ corresponding to the
transition dynamics depicted in Fig.~\ref{fig:DetailedBalancedMarkov} is given
in App. \ref{app:SpecificRatchetEnergetics}. Due to the repetitive nature of
the dynamics, the transducer reaches an asymptotic state
(App.~\ref{app:GeneralizedRatchetEnergetics}) such that its probability
distribution does not change from one interaction interval to another.

Now, consider the transducer's response when driven by a period-$2$ input
process. Appendix \ref{app:SpecificRatchetEnergetics} calculates the work and
entropy changes in the asymptotic limit, finding:
\begin{align}
\label{eq:Work2}
\langle W \rangle & = \frac{1 - \delta}{e} k_B T \ln{2} ~, \\
\Delta \H_1 & = 0 ~,~ \text{and} \\
\Delta \hmu & = \H\left( \frac{1-\delta}{e} \right)
  ~.
\label{eq:Expressions} 
\end{align} 
The work expression follows from the definition in Eq.~(\ref{eq:Work}). The
single-symbol entropy difference $\Delta \H_1$ vanishes since the output tape
consists of random, but still equal, mixtures of $0$'s and $1$'s, as did the
input tape. The entropy rate change $\Delta \hmu$, though, is generally
positive since, although the input entropy rate vanishes, the ratchet adds some
randomness to the output.

From Eq.~(\ref{eq:Expressions}), we have a clear violation of
Eq.~(\ref{eq:SecondLaw1}). Whereas, Eq.~(\ref{eq:SecondLaw2}) still holds:
\begin{align}
0 = \Delta \H_1 < \frac{\langle W \rangle}{k_B T \ln{2}} \leq \Delta h_\mu
  ~.
\label{eq:SecondLaw}
\end{align}
Since Ref. \cite{Boyd15a} established Eq.~(\ref{eq:SecondLaw2}) for all finite
ratchets, this difference in the bounds is expected. Nonetheless, it worth
calling out in light of recent discussions in the literature \cite{Mand15a}. In
any case, these results confirm the conclusion that to properly bound all
finite information ratchets, including memoryful ratchets driven by memoryful
inputs, we must use Eq.~(\ref{eq:SecondLaw2}) rather than
Eq.~(\ref{eq:SecondLaw1}).

\subsection{Dynamical Ergodicity and Synchronization}

\begin{figure}[tbp]
\centering
\includegraphics[width=.5\columnwidth]{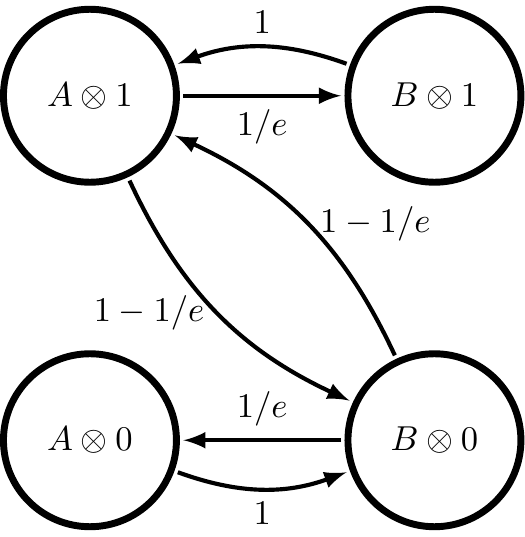}
\caption{\textbf{Ratchet dynamics absent the synchronizing state}: Assuming
	system parameters $\delta$ and $\gamma$ are set to zero, state $C$ becomes
	inaccessible and the ratchet's joint state-symbol dynamics become
	restricted to that shown here---a truncated form of the dynamics of Fig.
	~\ref{fig:DetailedBalancedMarkov}.
	}
\label{fig:PathDiagramBase}
\end{figure}

To provide intuition behind the work expression of Eq.~(\ref{eq:Work2}), let's
now analyze the ratchet's operation. This reveals a novel synchronization
mechanism that's responsible for nonzero work production. First, consider the
case in which the engine parameters $\delta$ and $\gamma$ are zero; that is,
the state $C$ is disconnected from $A$ and $B$. This effectively deletes $C$
from the joint dynamic, as shown in Fig. \ref{fig:PathDiagramBase}. This
restricted model has the topology considered in our previous work
~\cite{Boyd15a}.

\begin{figure}[tbp]
\centering
\includegraphics[width=1\columnwidth]{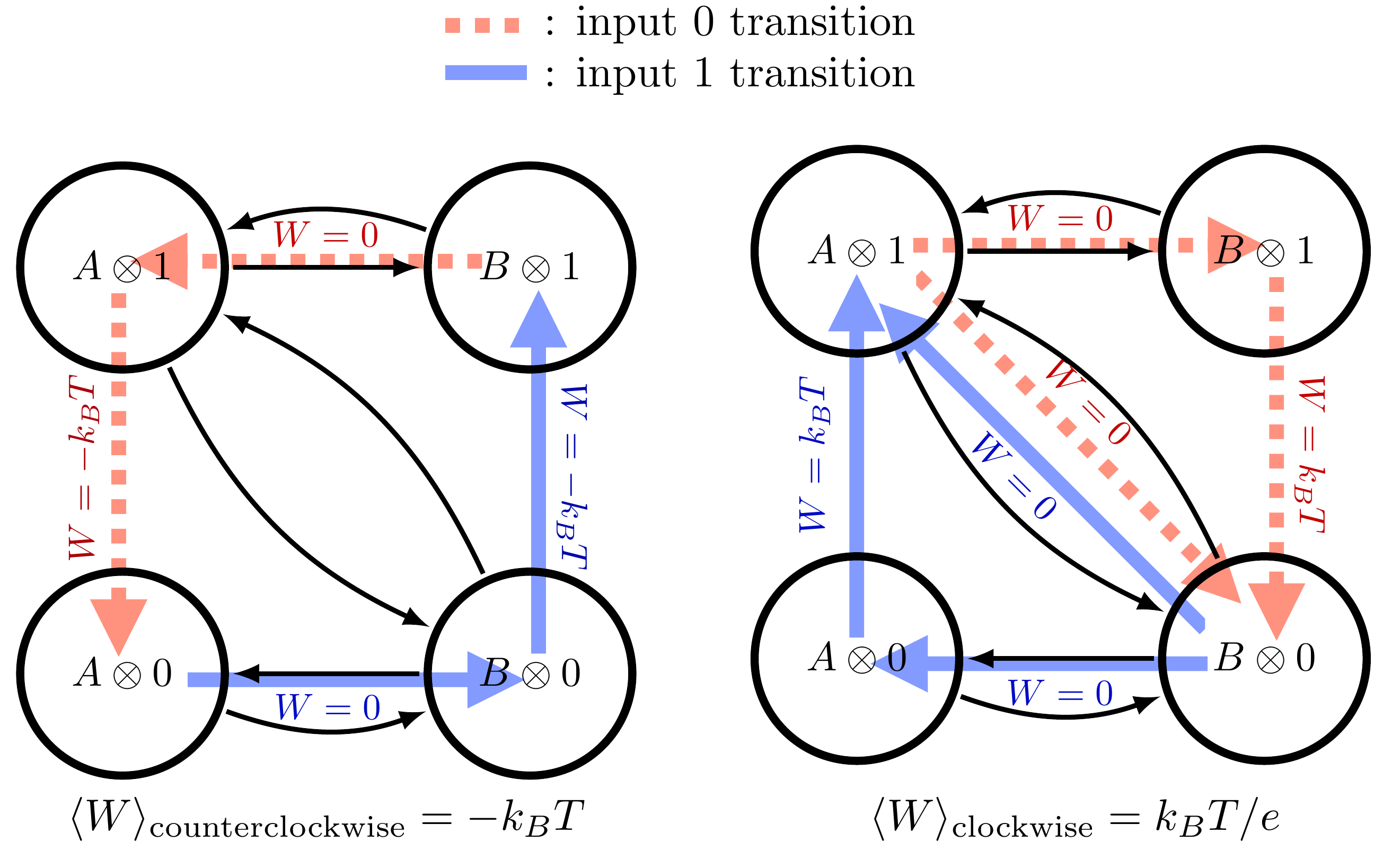}
\caption{\textbf{Two dynamical modes of the ratchet while driven by a
	period-$2$ input process:} (a) \emph{Counterclockwise} (left panel):
	ratchet is out of synchronization with the input tape and makes a steady
	counterclockwise rotation in the composite space of the Demon and the
	interacting cell. Work is steadily dissipated at the rate $-k_B T$ per pair
	of input symbols and no information is exchanged between the ratchet and
	the information reservoir. (b) \emph{Clockwise} (right panel): ratchet is
	synchronized with the input correlated symbols on the tape, information
	exchange is nonzero, and work is continually accumulated at the rate $k_B
	T/ e$ per pair of input symbols.
	}
\label{fig:PathDiagram}
\end{figure}

It turns out that the ratchet has two equally likely dynamical modes, let's
call them \emph{clockwise} and \emph{counterclockwise}. When in each mode, the
ratchet behavior is periodic in time. The modes are depicted in
Fig.~\ref{fig:PathDiagram}, with the counterclockwise mode on the left and the
clockwise mode on the right. The dashed (red) arrows show the paths taken
through the joint state space due to an interaction transition followed by a
switching transition when the switching transition is driven by input $0$. And,
the solid (blue) arrows show the paths taken when the switching transition is
driven by input $1$. The labels on the arrows indicate the amount of work done
in the associated transitions. The clockwise mode extracts $k_B T/e$ amount of
work per bit, while the counterclockwise mode expends $k_B T$ amount of work
per bit.

There is a simple way to understand the existence and work performance of the
two modes. Consider the counterclockwise mode first. The left
state-transition diagram in Fig.~\ref{fig:PathDiagram} shows this mode arises
when $A \otimes 0$ or $B \otimes 1$ happens to be the initial joint state.
First, there is a horizontal interaction transition to a lower energy state.
The energy difference $k_B T$ is fully dissipated in the thermal reservoir with
no exchange of energy with the work reservoir. Then, there is a vertical
switching transition to a higher-energy state. The required energy $k_B T$ is
taken from the work reservoir with no exchange of energy with the thermal
reservoir. This energy is then dissipated as heat in the thermal reservoir at
the next horizontal transition. The net amount of work produced per symbol---the
net amount of energy supplied to the work reservoir---is $\langle W \rangle = -
k_B T$.

Similarly, consider the clockwise mode. The righthand state-transition diagram
in Fig.~\ref{fig:PathDiagram} shows that this mode arises when $A \otimes 1$
or $B \otimes 0$ is the initial joint state. First, there is an interaction
transition along either the horizontal or diagonal paths of the Markov chain.
(The horizontal transitions are opposite to those of the counterclockwise
mode.) From microscopic reversibility, the horizontal interaction transitions
lead to $k_B T$ energy taken from the thermal reservoir in order to move into
higher energy states. No energy is exchanged with the work reservoir. On the
diagonal transitions, on the other hand, no energy is exchanged with either
reservoir.  Then, there is a switching transition, which corresponds to a
vertical transition to a lower energy state if the horizontal interaction
transition was made just before. The energy difference $k_B T$ is given to the
work reservoir.  However, if the diagonal transition was made, then the
switching transition does not change the state and there is no work done. As
shown in the figure, there are two possible paths the system can take between
$A \otimes 1$ and $B \otimes 0$ in one operation cycle of the ratchet: $\{A
\otimes 1 \rightarrow B \otimes 1 \rightarrow B \otimes 0\}$ and $\{A \otimes 1
\rightarrow B \otimes 0 \rightarrow B \otimes 0\}$.   The same is true of
transitions from $B \otimes 0$ to $A \otimes 1$: $\{B \otimes 0 \rightarrow A
\otimes 0 \rightarrow A \otimes 1\}$ and $\{B \otimes 0 \rightarrow A \otimes 1
\rightarrow A \otimes 1\}$.  Averaging over the probabilities of the two
fundamental paths, the net average work produced is $\langle W \rangle = k_B T
/e$. (See App.  \ref{app:SpecificRatchetEnergetics} for details.)

If the initial ratchet state is uncorrelated with the input HMM state, the
clockwise and the counterclockwise modes occur with equal probability. Once in
a particular mode, the ratchet cannot switch over to the other mode. In this
sense, the two modes act as two different attractors for the Demon's
joint state-symbol dynamics. In other words, the system is dynamically
nonergodic, leading to nonergodic work production: either time averaged $-k_B
T$ or $k_BT/e$. In this case, the ratchet dissipates on average $k_B T
(1-1/e)/2 $ units of energy from the work reservoir into the thermal
reservoir as heat.

Comparing this ergodic ratchet, in which nonergodicity plays a dynamic and
transient role, to the nonergodic engine discussed earlier is in order.
Nonergodic engines (those driven by nonergodic input processes) can exhibit
functional behavior when averaged over an ensemble of input realizations. As
shown in Ref. \cite{Chap15a}, Maxwell's refrigerator \cite{Mand2013} can
refrigerate when driven by the nonergodic process consisting of two infinitely
long realizations, one of all $0$s and the other of all $1$s. Similar to our
ratchet driven by (ergodic) period-$2$ sequences, the refrigerator has two
principle modes: the ratchet is driven by all $0$s and refrigerates versus the
ratchet is driven by all $1$s and dissipates. However, one of these two modes is chosen
at random in the beginning of a ratchet trial and remains fixed. This yields
refrigeration that differs from the ensemble average (over the nonergodic input
realizations).  However, we can achieve robust and functional work production
in our period-2 ratchet, by coupling modalities dynamically via the $C$ state.
Then, on every trial, the engine functions.

Let's explain how its emergent nonergodicity makes this function robust. For
$\delta \neq 0$ and $\gamma \neq 0$, state $C$ becomes accessible to the
ratchet, changing the stability of the counterclockwise attractor. And, this
allows positive work production. (From here on we consider the original, full
ratchet in Fig.~\ref{fig:DetailedBalancedMarkov}.) We make a heuristic argument
as to why the ratchet can generate positive net work using state $C$.

\begin{figure}
\centering
\includegraphics[width=1\columnwidth]{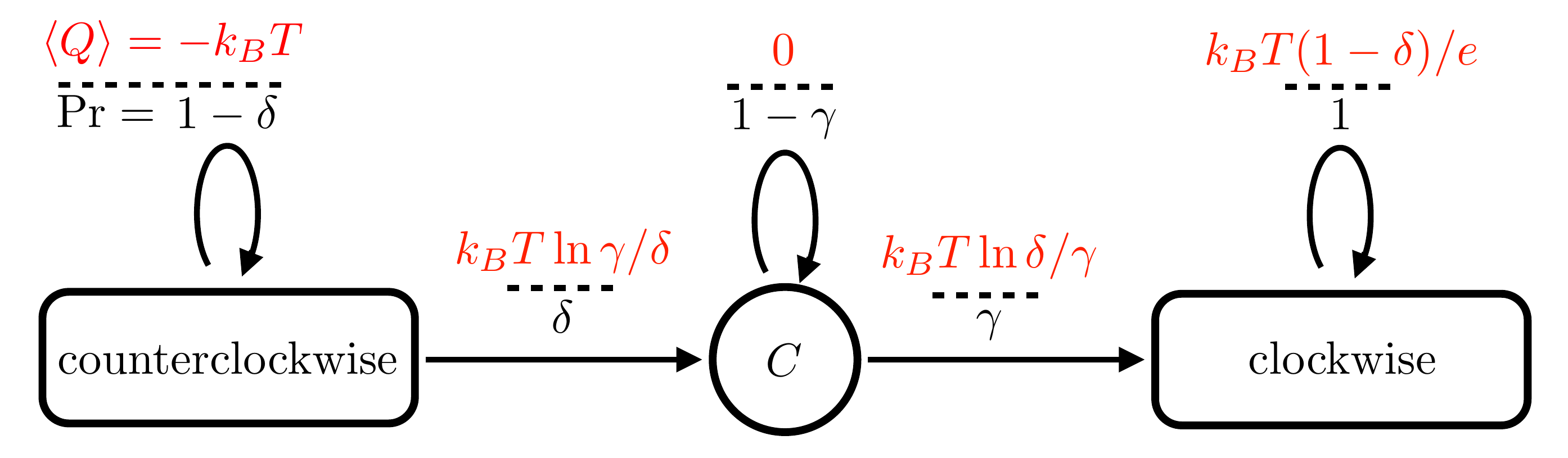}
\caption{\textbf{Crossover from the dissipative, counterclockwise mode to
   the generative, clockwise mode via synchronizing state $C$:} Even though
   the microscopic dynamics satisfy time-reversal symmetry, a crossover is possible only from the counterclockwise mode to the clockwise mode because of the topology of the joint state space.  With transitions between $C$ and the other two modes, the engine becomes ergodic among its dynamic modes.  The heats and transition probabilities are shown above each arrow.
	}
\label{fig:Synchronizing}
\end{figure}

$C$'s addition creates a ``path" for the ratchet to shift from the dissipative,
counterclockwise mode to the generative, clockwise mode. And, the latter
becomes the only attractor in the system. In other words, the counterclockwise
dynamical mode becomes a purely transient mode and the system becomes
dynamically ergodic. The situation is schematically shown in
Fig.~\ref{fig:Synchronizing}, where the arrows denote allowed transitions in
the dynamical sense. Heat and probability values of the transitions are shown
there along each arrow.  Recall that in the counterclockwise mode, the joint state 
is either $A \otimes 0$ or $B \otimes 1$ at the beginning of each interaction interval.  
According to the Markov model, both these states have probability $\delta$ of transitioning 
to a $C$ state during interaction transitions.  Thus, as depicted, $\delta$ is the probability of
transitioning from the counterclockwise mode to $C$.

Once in state $C$, the ratchet cannot return to the counterclockwise mode,
despite the fact there is probability $\gamma$ of transitioning back to either
$A \otimes 0$ or $B \otimes 1$ in an interaction transition. This is because
the following switching transition immediately changes $A \otimes 0$ to $A
\otimes 1$ and $B \otimes 1$ to $B \otimes 0$. That is, the system is in the
clockwise mode at the beginning of the next interaction interval. Thus, with
probability $\gamma$ the system makes a transition to the clockwise mode. After
this transition, the system is necessarily synchronized, and it is impossible
to transition out of the synchronized dynamic. In this way, the ratchet
asymptotically extracts a positive amount of average heat from the environment,
$\langle Q \rangle= k_B T (1-\delta)/e$ per symbol. Asymptotic heat extraction
is the same as the work production for finite ratchets, confirming
Eq.~(\ref{eq:Work2}).  Since the ratchet must move through $C$ to arrive at the
recurrent, clockwise, work-producing dynamic, we decide to start the ratchet in
$C$. $C$ serves as a synchronization state in that it is necessary for the
ratchet state to synchronize to the input tape: once the ratchet transitions
out of the $C$ state, its internal states are synchronized with the input HMM
states such that it produces work.

\subsection{Trading-off Work Production Against Synchronization Rate and Work}

With Fig.~\ref{fig:Synchronizing} in mind, we can define and calculate several
quantities that are central to understanding the ratchet's thermodynamic
functionality as functions of its parameters $\delta$ and $\gamma$: the
\emph{synchronization rate} $R_\text{sync}$ and the \emph{synchronization heat}
$Q_\text{sync}$ absorbed during synchronization. $R_\text{sync}$ is the inverse
of the average number of time steps until transitioning into the clockwise
mode. It simplifies to the probability $\gamma$ of transitioning into the
clockwise mode:
\begin{align*}
R_\text{sync}(\delta,\gamma)&=\frac{1}{\langle t/\tau \rangle} \\
  & = \frac{1}{\gamma \sum_{i=0}^\infty(i+1) (1-\gamma)^i} \\
  & = \gamma
  ~.
\end{align*}
The heat $Q_\text{sync}$ absorbed when synchronizing is the change in energy of
the joint state as the ratchet goes from the synchronizing states ($C\otimes 0$
or $C \otimes 1$) into the recurrent synchronized states ($A \otimes 0$ or $B
\otimes 1$):
\begin{align*}
Q_\text{sync}(\delta, \gamma) & = k_B T \ln \frac{\delta}{\gamma}
  ~.
\end{align*}
This is minus the energy dissipation required for synchronization.

\begin{figure}[tbp]
\centering
\includegraphics[trim = 0 0 0 0, width=0.9\columnwidth]{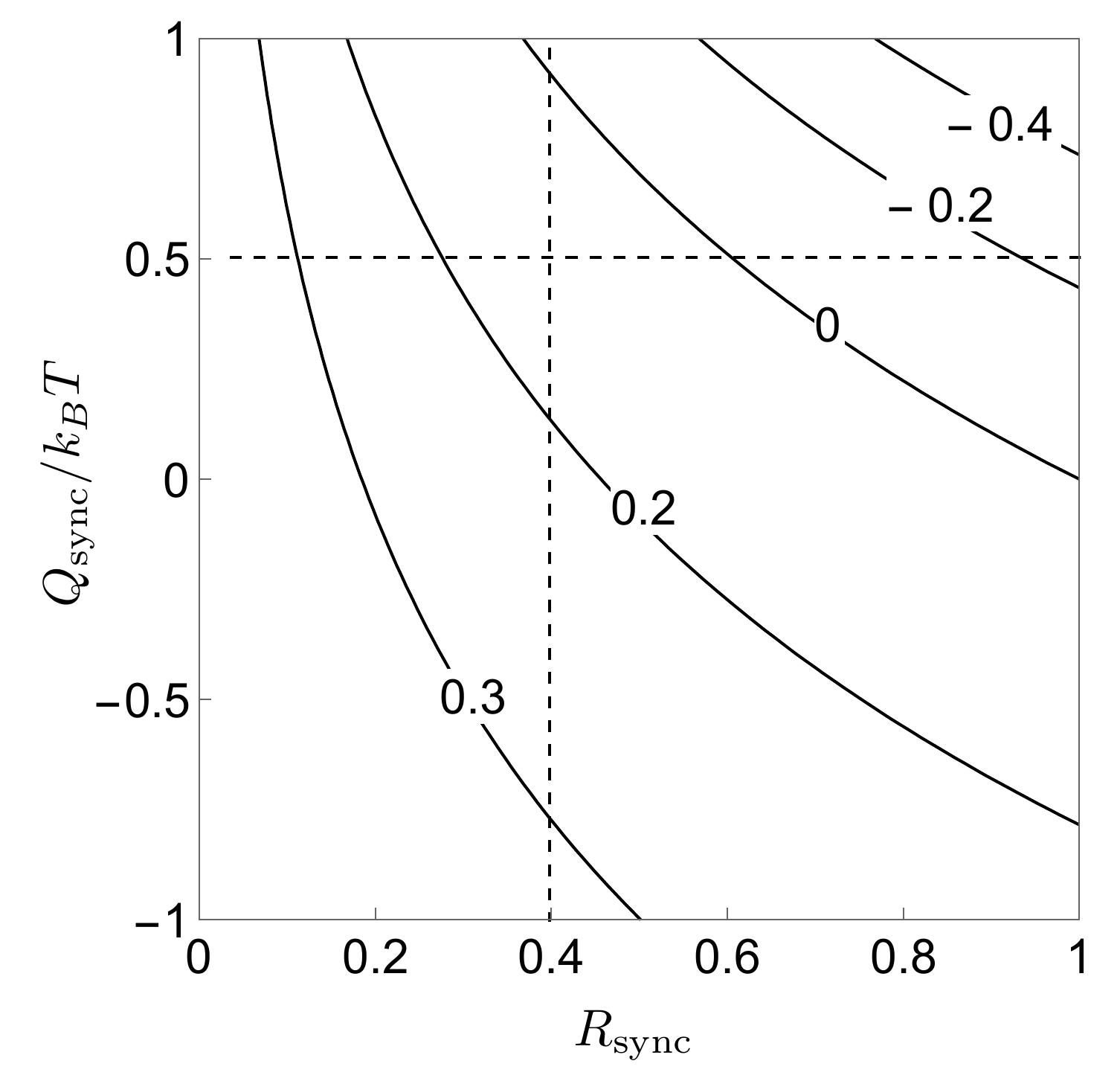}
\caption{\textbf{Trade-off between average work production, synchronization
	rate, and synchronization heat:} Contour plot of average extracted work per
	symbol $\langle W \rangle $ as a function of rate of synchronization
	$R_\text{sync}$ and synchronization heat $Q_\text{sync}$ using
	Eq.~(\ref{eq:Tradeoff}). Work values are in the unit of $k_B T$. Numbers
	labeling contours denote the average extracted work $\langle W \rangle$. If
	we focus on any particular contour, increasing $R_\text{sync}$ leads to a
	decrease in $Q_\text{sync}$ and vice versa. Similarly, restricting to a
	fixed vale of $R_\text{sync}$, say the vertical $R_\text{sync} = 0.4$ line,
	increasing $Q_\text{sync}$ decreases values of $\langle W \rangle $.
	Restricting to a fixed vale of $Q_\text{sync}$, say the horizontal
	$Q_\text{sync} = 0.5$ line, increasing $R_\text{sync}$ going to the right
	also decreases $\langle W \rangle $.
	}
\label{fig:Tradeoff}
\end{figure}

Much like the speed, energy cost, and fidelity of a computation~\cite{Benn79,
Muru12, Zulk14a, Lahi16a}, these two quantities and the average extracted work
per symbol obey a three-way tradeoff in which each pair is inversely related,
when holding the third constant. This is expressed most directly by combining
the expressions above into a single relation that is independent of $\delta$
and $\gamma$:
\begin{align}
Q_\text{sync}+k_B T \ln R_\text{sync} -k_B T
  \ln \left( 1-\frac{e \langle W \rangle }{k_B T}\right) = 0
  ~.
\label{eq:Tradeoff}
\end{align}
Figure~\ref{fig:Tradeoff} illustrates this trade-off. Analytically, the same
interdependence appears when taking the partial derivatives of the quantities
with respect to each other:
\begin{align*}
\frac{\partial Q_\text{sync}}{\partial \langle W \rangle }
  & = -\frac{-k_B T e}{k_B T- e \langle W \rangle} ~, \\
\frac{\partial Q_\text{sync}}{\partial R_\text{sync} }
  & = -\frac{-k_B T }{R_\text{sync}} ~, ~\text{and}\\
\frac{\partial \langle W \rangle }{\partial R_\text{sync} }
  & = -\frac{k_B T-e \langle W \rangle }{e R_\text{sync}}
  ~.
\end{align*}
These all turn out to be negative over the physical range of parameters:
$\langle W \rangle \in (-\infty, 1/e]$, $R_\text{sync} \in [0, 1]$, and
$Q_\text{sync} \in (-\infty, \infty)$.

The ratchet's successful functioning derives from the fact that it exhibits a
dynamical mode that ``resonates" with the input process correlation in terms
of work production and that this mode can be made the only dynamical
attractor. In other words, an essential element in constructing our ratchet its
ability to synchronize its internal states with the effective states of the
input process. This appears to be a basic principle for leveraging memoryful
input processes and, more generally, correlated environments.

\begin{figure}[tbp]
\centering
\includegraphics[width=.9\columnwidth]{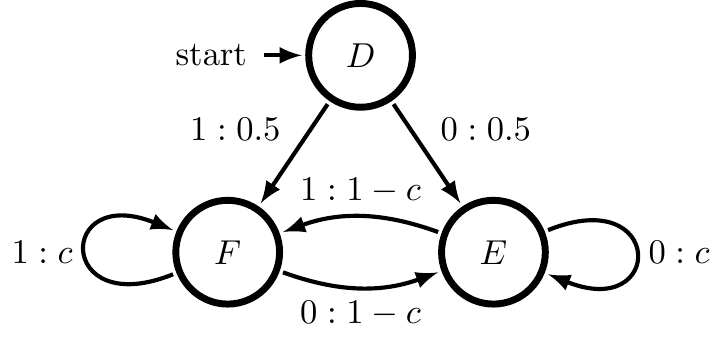}
\caption{\textbf{Noisy phase-slip period-$2$ (NPSP2) process:} As with the
	exact period-$2$ process of Fig.~\ref{fig:Period2}, its HMM has a transient
	start state $D$ and two recurrent causal states $E$ and $F$.  Starting from
	$D$, the process makes a transition either to $E$ or $F$ with equal
	probabilities while outputting $0$ or $1$, respectively. Once in state $E$,
	the process either stays with probability $c$ and outputs a $0$ or makes a
	transition to state $F$ with probability $1-c$ and outputs a $1$. If in
	state $F$, the process either stays with probability $c$ and outputs a $1$
	or makes a transition to state $E$ with probability $1-c$ and outputs a
	$0$. For small nonzero $c$, the output is no longer a pure alternating
	sequence of $0$s and $1$s, but instead randomly breaks the period-$2$
	phase. For $c = 1/2$, the generated sequences are flips of a fair coin.
	The process reduces to that in Fig.~\ref{fig:Period2}, if $c = 0$.
	}
\label{fig:PhaseSlips}
\end{figure}

\section{Fluctuating Correlated Environments}
\label{sec:CorruptInput}

The preceding development considered a perfectly correlated environment that
generates an input to the ratchet in which a $0$ is always followed by $1$ and
a $1$ by $0$. Of course, this is an artificial and constrained input. It's
purpose, though, was to isolate the role of structured, correlated environment
signals and how a thermodynamic ratchet can leverage that order to function as
an engine. Practically, though, it is hard to come by such perfectly correlated
sequences in Nature. One expects sequences to involve errors, say where a $0$
is sometimes followed by a $0$ and a $1$ by $1$. Such \emph{phase slips} are
one kind of error with which a thermodynamically functioning ratchet must
contend.

In particular, whenever a phase slip occurs the ratchet is thrown out of its
synchronization with the input, possibly into the dissipative, counterclockwise
dynamical mode. Due to the presence of the synchronizing mechanism, shown in
Fig.~\ref{fig:Synchronizing}, the ratchet can recover via transiting through
the synchronizing state $C$. If the frequency of phase slips is sufficiently
low, then, the ratchet can still produce work, only at a lower rate. If the
phase slip frequency is high enough, however, the ratchet does not have
sufficient time in the clockwise mode to recover the work lost in the
counterclockwise mode before it relaxed to the clockwise mode. At this error
level the ratchet stops producing work; it dissipates work even on average.
This suggests there is a critical level of input errors where a transition from
a functional to nonfunctional ratchet occurs. This section analyzes the
transition, giving an exact expression for the critical phase-slip frequency at
which the ratchet stops producing work.

To explore the ratchet's response to such errors, we introduce phase slips into
the original period-$2$ input process. They occur with a probability $c$,
meaning that after every transition, there is a probability $c$ of emitting the
same symbol again and remaining in the same hidden state rather than emitting
the opposite symbol and transitioning to the next hidden state. An HMM
corresponding this period-$2$ phase-slip dynamics is shown in
Fig.~\ref{fig:PhaseSlips}---the \emph{noisy phase-slip period-$2$} (NPSP2)
process. It reduces to the original, exactly periodic process generated by the
HMM in Fig. \ref{fig:Period2} when $c = 0$.

It is now straightforward to drive the ratchet
(Fig.~\ref{fig:DetailedBalancedMarkov}) inputs with the NPSP2 process
(Fig.~\ref{fig:PhaseSlips}) and calculate exactly the average work production
per symbol using Eq.~(\ref{eq:Work}). Appendix \ref{app:SpecificRatchetEnergetics}
does this for all values of $\delta$, $\gamma$, and $c$. Here, let's first
consider the special case of $\gamma=1$. This is the regime in which the
ratchet is most functional as an engine since, if the ratchet produces positive
work, then $\gamma=1$ maximizes that work production.

With $\gamma=1$, once the ratchet is in state $C$, it immediately synchronizes
in the next interaction interval. In this case, $\delta$ parametrizes the
relationship between the average work done when synchronized and the rate of
synchronization. The higher $\delta$ is, the less work the ratchet extracts
while synchronized, but the more often it transitions to the synchronizing
state---recall Fig.~\ref{fig:Synchronizing}---allowing it to recover from phase
slips. The calculation for $\gamma=1$ yields an average work rate (App.
\ref{app:SpecificRatchetEnergetics}):
\begin{align}
\langle W \rangle (\delta, c)
  = \frac{(1-\delta)[\delta+c-c(2\delta +e)]}{2ec + \delta e (1-c)}
  ~.
\label{eq:Work3}
\end{align}
Thus, over the whole parameter space $c,\delta \in [0,1]$, the average work 
varies over the range:
\begin{align*}
\langle W \rangle (\delta, c) \in \frac{k_B T}{e}
  \left[-\frac{e-1}{2},1\right]
  ~.
\end{align*} 

\begin{figure}[tbp]
\centering
\includegraphics[width=1.0\columnwidth]{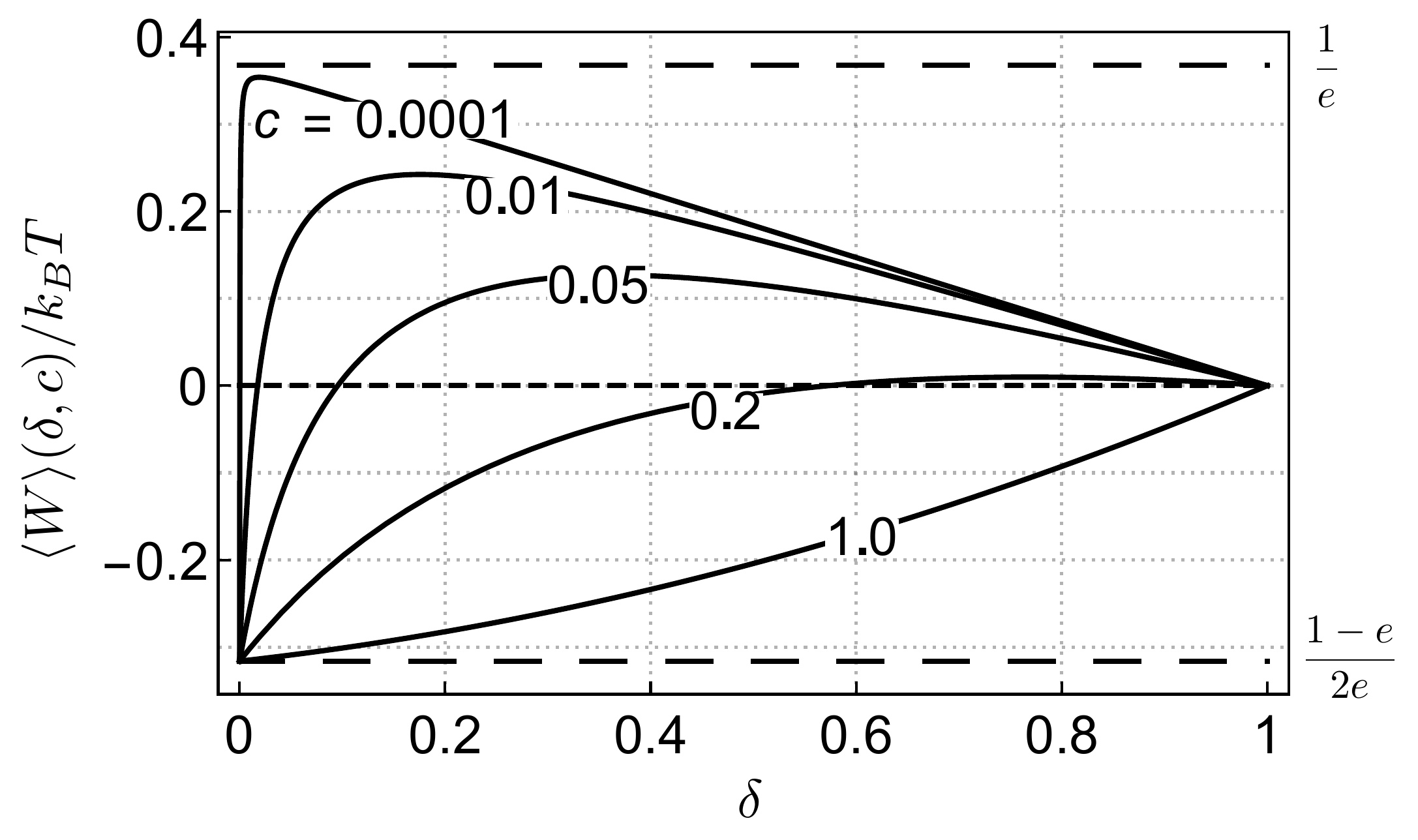}
\caption{\textbf{Average work production per symbol versus synchronization rate
	$\delta$ and phase slip rate $c$ at fixed $\gamma = 1$.} Labels on the
	curves give $c$ values. Long-dashed lines give the upper and lower bounds
	on work production: It is bounded above by $k_B T / e$ and below by
	$k_B T (1-e) / (2e)$. (See text.)
	}
\label{fig:ErrorCorrectionWork}
\end{figure}

Figure~\ref{fig:ErrorCorrectionWork} shows how the work production varies with
$\delta$ for different values of $c$. No matter the value of $c$, at $\delta =
0$ the average work attains its lower limit of $- k_B T (e - 1)/2 e$, which is
the average work produced when both clockwise and counterclockwise modes have
equal probability. As $\delta$ increases, there is an increase in the the
average work until it reaches $0$ at a particular value $\delta^*(c)$. Below
$\delta^*(c)$---i.e., within the range $0 \leq \delta \leq\delta^*(c)$---the
system consumes work; whereas above $\delta^*(c)$, the system acts as an
engine, producing net positive work. Figure~\ref{fig:ErrorCorrectionWork} shows
that $\delta^*(c)$ is an increasing function of $c$, starting with $0^+$ as $c$
tends to $0$ and ending up at $1$ as $c$ tends to unity.

The dependence is nonlinear, with sharp changes near $c = 0$ and saturating
near $c = 1$. Since the average work vanishes as $\delta$ tends to $1$
independent of $c$, there is a value of $\delta_\text{max}(c)$ where the
engine's work production is maximum. This maximum work $W_\text{max}(c)$ is
closer to its upper limit $k_B T / e$ for smaller values of $c$. As we increase
$c$, there is a decrease in $W_\text{max}(c)$ until it vanishes at $\delta =
1$.

\begin{figure}[tbp]
\centering
\includegraphics[width=\columnwidth]{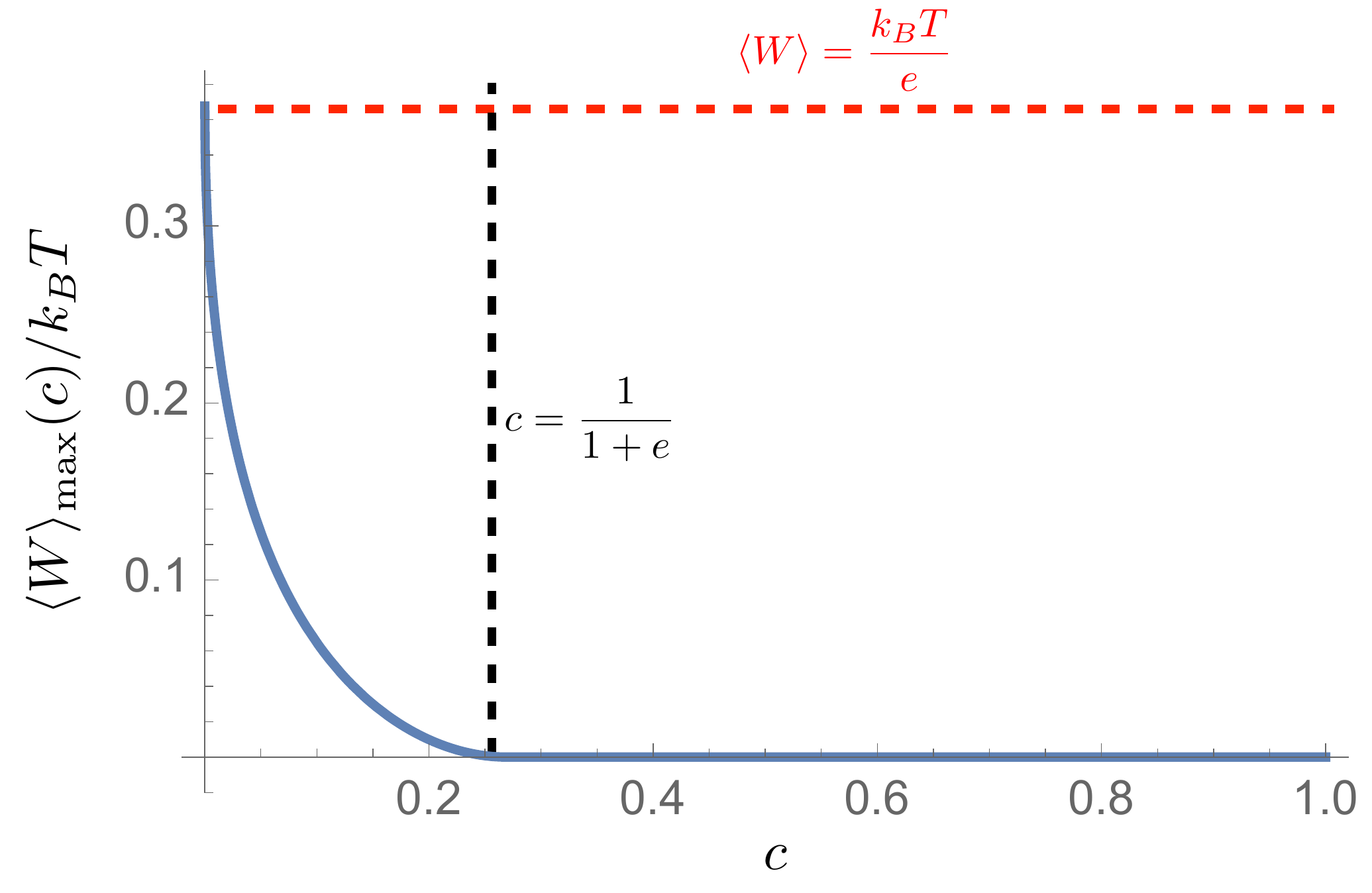}
\caption{\textbf{Maximum work production versus phase-slip rate $c$.}
	Maximum work production decreases with $c$ from $k_B T/e$ at $c=0$
	to $0$ when $c\geq c^* =  1/(1+e)$.
	}
\label{fig:ErrorCorrectionMaxWork}
\end{figure}

Figure~\ref{fig:ErrorCorrectionMaxWork} shows the dependence of
$W_\text{max}(c)$ as a function of error rate $c$, revealing a critical value
$c^* = 1/1+e$ beyond which $W_\text{max}$ vanishes. Thus, if the phase-slip
frequency is too high, the ratchet cannot produce net positive work regardless of how quickly it synchronizes.  This special value $c^*$ actually partitions the expressions for $\delta^*(c)$,
$\delta_\text{max}(c)$, and $W_\text{max}(c)$ into piecewise functions:  
\begin{align}
\label{eqn:Deltas}
\delta^*(c) & = \begin{cases}
\frac{(1-e)c}{2c-1} &\text{ if } c\leq \frac{1}{1+e} 
\\  1 &\text{ if }c>\frac{1}{1+e}
\end{cases} \\
\delta_\text{max}(c) & = \begin{cases}
 \frac{4 c^2+\alpha-2 c}{2 c^2-3 c+1} &\text{ if } c\leq \frac{1}{1+e} 
 \\ 1& \text{ if }c>\frac{1}{1+e} 
 \end{cases}\\
W_\text{max}(c) & = \begin{cases}
k_B T \frac{-2 \alpha +c (e-(5+e) c)+1}{e (c-1)^2} &\text{ if } c\leq \frac{1}{1+e} 
\\ 0 &\text{ if }c>\frac{1}{1+e} 
\end{cases}
  ,
\end{align}
where $\alpha = \sqrt{c \left(2 c^2+c-1\right) ((3+e) c-e-1)}$.

The results in Fig.~\ref{fig:ErrorCorrectionMaxWork} should not be too broadly
applied. They do not imply that positive net work cannot be extracted for the
case $c > c^*$ for \emph{any} information ratchet. On the contrary, there exist
alternatively designed ratchets that can extract positive work even at $c = 1$.
However, the design of such ratchets differs substantially from the current
one. Sequels will take up the task of designing and analyzing this broader
class of information engines.

\begin{figure}[tbp]
\centering
\includegraphics[width=\columnwidth]{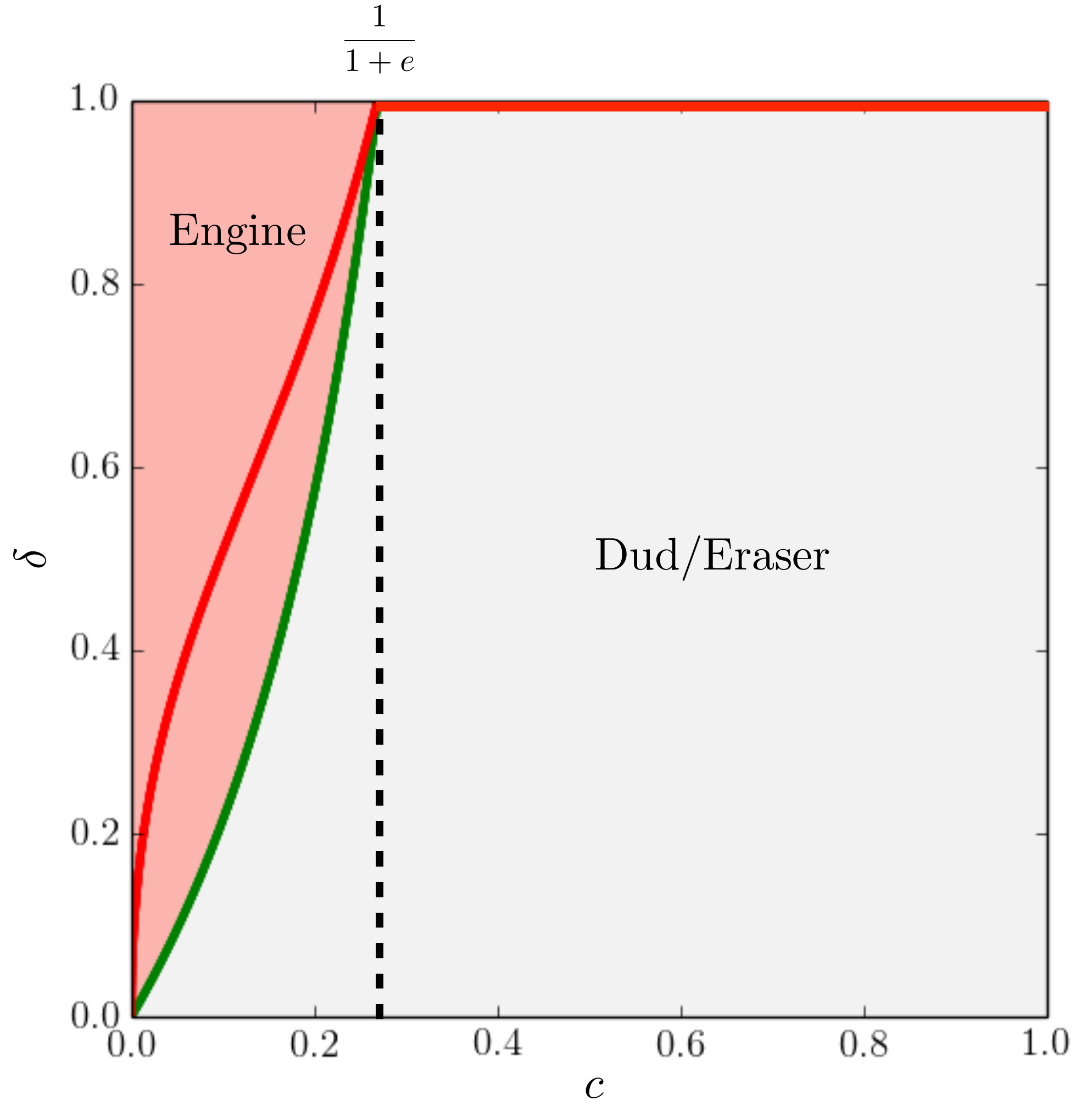}
\caption{\textbf{Ratchet thermodynamic-function phase diagram:} In the leftmost
  (red) region, the ratchet behaves as an engine, producing positive work. In
  the lower right (gray) region, the ratchet behaves as either an eraser
  (dissipating work to erase information) or a dud (dissipating energy without
  erasing information). The solid (red) line indicates the parameter values
  that maximize the work output in the engine mode. The dashed (black) line
  indicates the critical value of $c$ above which the ratchet cannot act as an
  engine.
  }
\label{fig:PhaseDiagram}
\end{figure}

Figure~\ref{fig:PhaseDiagram} combines the results in
Figs.~\ref{fig:ErrorCorrectionWork} and~\ref{fig:ErrorCorrectionMaxWork} into
phase diagram summarizing the ratchet's thermodynamic functionality. It
illustrates how the $\delta$-$c$ parameter space splits into two regions:  the
leftmost (red) region where the ratchet produces work, behaving as an engine,
and the lower right (gray) region where the ratchet consumes work, behaving as
either an information eraser (using work to erase information in the bit
string) or a dud (dissipating work without any erasure of information). It also
shows that $c^* = 1 / (1+e)$ corresponds to both the point at which
$\delta_\text{max}(c)$ reaches $1$ and the point at which it is no longer
possible to extract work from the input, independent $\delta$. This is the
point where phase slips happen so often that the ratchet finds it impossible to
synchronize for long enough to extract any work.

\section{Conclusion}

We extended the functionality of autonomous Maxwellian Demons by introducing a
new design for information engines that is capable of extracting work purely
out of temporal correlations in an information source, characterized by an
input HMM. This is in marked contrast with previous designs that can only
leverage a statistically biased information source or the mutual, instantaneous
correlation between a pair of information sources \cite{Mand012a, Mand2013,
Bara2013, Boyd15a, Merh15a}. Our new design is especially appropriate for
actual physical construction of information engines since physical, chemical,
and biological environments (information sources) almost always produce
temporally correlated signals.

The new design was inspired by trying to resolve conflicting bounds on the work
production for information engines. On the one hand, Eq.~(\ref{eq:SecondLaw1})
for monitoring information content only of isolated symbols suggests that no
work can be produced from temporal correlations in input string; whereas, on
the other, using entropy rates Eq.~(\ref{eq:SecondLaw2}) indicates these
correlations are an excellent resource. We showed, in effect, that this latter
kind of correlational information is a thermodynamic fuel.

To disambiguate the two bounds, we described the exact analytical procedure to
calculate the average work production for an arbitrary memoryful channel and a
HMM input process. The result is that it is now abundantly clear which bounds
hold for correlated input processes.

We considered the specific example of a period-$2$ process for the input tape
(Fig.~\ref{fig:Period2}), since it has structure in its temporal correlations,
but no usable single-symbol information content. The ratchet we introduced to
leverage this input process requires three memory states
(Fig.~\ref{fig:DetailedBalancedMarkov}) to produce positive work. This
memoryful ratchet with a memoryful input process violates
Eq.~(\ref{eq:SecondLaw1}), establishing Eq.~(\ref{eq:SecondLaw2}) as the proper
information processing Second Law of thermodynamics.

It is intuitively appealing to think that ratchet memory must be in consonance
with the input process' memory to generate positive work. In other words, the
ratchet must be memoryful and be able to synchronize itself to the structured
memory of the input HMM to be functional. We confirmed that this is indeed the
case in general with our expression for work. If the ratchet has no memory, the
only ``structure'' of consequence in the input process is simply, provably, the
isolated-symbol statistical bias.

We see this nascent principle more concretely in the operation of the ratchet
as it responds to the period-$2$ process. Critical to its behaving as an engine
is the presence of state $C$ (Fig.~\ref{fig:DetailedBalancedMarkov}) through
which the ratchet synchronizes itself to the input. As shown in
Fig.~\ref{fig:Synchronizing}, the synchronizing state $C$ allows the system to
make an irreversible transition from the counterclockwise, dissipative mode
into the generative, clockwise mode. It demonstrates how key it is that the
ratchet's effective memory match that of the input process generator.

We also discovered an intriguing three-way tradeoff (Fig.~\ref{fig:Tradeoff})
between synchronization rate, synchronization heat (that absorbed during
synchronization), and asymptotic average work production. For example, if the
Demon keeps the synchronization rate fixed and increases the synchronization
heat, there is a decrease in the average work production. In other words, if
the Demon becomes greedy and tries to extract energy from the thermal reservoir
even during synchronization, on the one hand, it is left with less work in the
end. If, on the other hand, the Demon actually supplies heat during the
synchronization step, it gains more work in the end! Similarly, if it keeps the
synchronization heat fixed, a slower rate of synchronization is actually better
for the average work production. If the Demon waits longer for the ratchet to
synchronize with its environment, it is rewarded more in terms of the work
production. Thus, the Demon is better off in terms of work, by being patient
and actually supplying more energy during synchronization. This three-way
tradeoff reminds one of a recently reported tradeoff between the rate, energy
production, and fidelity of a computation \cite{Zulk14a}.

We then considered the robustness of our design in a setting in which the input
process is not perfectly periodic, but has random phase slips
(Fig.~\ref{fig:PhaseSlips}). As a result, the dissipative regime is no longer
strictly transient. Every so often, the ratchet is thrown into the dissipative
regime induced by the phase slips, after which the ratchet attempts to
resynchronize to the generative mode. Thus, the ratchet seems remarkably robust
with respect to the phase-slip errors, being able to dynamically correct its
estimation of the input's hidden state due to the synchronization mechanism.
This is true, however, only up to a certain probability of phase slips, beyond
which the dissipative regime is simply too frequent for the ratchet to generate
any work. For the region in which the ratchet is capable of generating work, we
found the parametric combination for its optimal functionality for a given
probability of phase slips (Fig.~\ref{fig:ErrorCorrectionWork}). We also
determined the maximum net work that the ratchet can produce
(Fig.~\ref{fig:ErrorCorrectionMaxWork}). Finally, we gave a phase diagram of
the ratchet's thermodynamic functionality over the control parameter space
formed by $\delta$ and $c$ for $\gamma = 1$ (Fig.~\ref{fig:PhaseDiagram}).

In this way, we extended the design of information engines to include memoryful
input processes and memoryful ratchets. The study suggests, via synchronization
and dynamical self-correction, there are general principles that determine how
autonomous devices and organisms can leverage arbitrary structure in their
environments to extract thermodynamic benefits.

Physical systems that demonstrate the thermodynamic equivalent of information
processing are by now numerous. Most, in contrast to the present design,
restrict themselves to single-step information processing. Moreover, many only
consider information processing comprising the erasure of a single bit, staying
within the setting of Landauer's Principle. The information-processing
equivalence principle strongly suggests a much wider set of computational
possibilities that use the capacity of stored information as a thermodynamic
resource.

Practically implementing an information engine on the nanoscale, say, will
require delicate control over system and materials properties. To achieve this
in a convincing way will demand an unprecedented ability to measure heat and
work. This has become possible only recently using single-electron
devices~\cite{Kosk15}, nanoelectronic mechanical systems (NEMS)~\cite{Kara09,
Math14}, and Bose-Einstein Condensates (BECs) \cite{ Ande95a,Davi95a,Brad95a}.
The results and methods outlined here go some distance to realizing these
possibilities by pointing to designs that are functionally robust and
resilient, by identifying efficient information engines and diagnosing their
operation, and by giving exact analytical methods for the quantitative
predictions necessary for implementation.

\section*{Acknowledgments}

The authors thank the Telluride Science Research Center for its hospitality
during this work's completion. As an External Faculty member, JPC thanks the
Santa Fe Institute for its hospitality during visits. This work was supported
in part by the U. S. Army Research Laboratory and the U. S. Army Research
Office under contracts W911NF-13-1-0390 and W911NF-12-1-0234.

\appendix

\section{Ratchet Energetics: General Treatment}
\label{app:GeneralizedRatchetEnergetics}

Here, we lay out the detailed calculations of the thermodynamic contributions
made by the ratchet's transducer and the environmental input process.

\subsection{Transducer Thermodynamic Contributions}

We consider the case where the ratchet exchanges energy only with the work
reservoir during the switching transitions and only with the heat reservoir
during the interaction transitions. During the $N$-th switching transition, the
ratchet ``exhausts" the $N$-th input bit $Y_N$ as the $N$-th output bit $Y'_N$
and couples with the input bit $Y_{N+1}$. The joint state of the ratchet and the
interacting bit changes from $X_{N+1}\otimes Y'_N$ to $X_{N+1}\otimes Y_{N+1}$.
The corresponding decrease in energy is supplied to the work reservoir. So, the
work output at the $N$-th switching transition $W_N$ is given by:
\begin{align}
W_N=E_{x_{N+1} \otimes y'_N}-E_{x_{N+1}\otimes y_{N+1}}
  ~,
\label{eq:W_N}
\end{align}
where $E_{x \otimes y}$ denotes the energy of the joint state $x \otimes y$. 
Via a similar argument, we write the heat absorbed by the ratchet during the $N$-th interaction
transition $Q_N$:
\begin{align*}
Q_N = E_{x_{N+1} \otimes y'_N}-E_{x_N \otimes y_N}
  ~.
\end{align*}

The main interest is in determining the asymptotic rate of work production:
\begin{align}
\langle W \rangle & = \lim_{N \rightarrow \infty} W_N \Pr(W_N)
  \nonumber \\ 
  & = \lim_{N \rightarrow \infty}
  \sum_{\substack{x_{N+1}, \\ y_{N+1},y'_N}}
  (E_{x_{N+1} \otimes y'_N}-E_{x_{N+1}\otimes y_{N+1}})
\label{eq:WAsymptotic3} \\
  & \quad\quad \times \Pr(X_{N+1}=x_{N+1},Y_{N+1}=y_{N+1},Y'_N=y'_N)
  \nonumber \\
  & = \sum_{x',y'}E_{x'\otimes y'}
  \lim_{N \rightarrow \infty} \Pr(X_{N+1}=x',Y'_{N}=y')
  \nonumber \\
  & \quad\quad - \sum_{x,y} E_{x\otimes y} \lim_{N \rightarrow \infty}
  \Pr(X_{N+1}=x,Y_{N+1}=y)
  \nonumber
  ~,
\end{align}
where the second line uses Eq.~(\ref{eq:W_N}) and the third relabels the
realizations in the sum $x$ and $x'$, since these are dummy variables in
separate sums.

Assuming the stationary distribution over the input variable and ratchet
variable exists, the asymptotic probability $\lim_{N \rightarrow \infty}
\Pr(X_{N+1}=x,Y_{N+1}=y)$ is the same as the asymptotic probability $\lim_{N
\rightarrow \infty} \Pr(X_{N}=x,Y_{N}=y)$, which was defined as
$\pi_{x\otimes y}$. In addition, note that the Markov matrix $M$ controlling
the joint ratchet-bit dynamic is stochastic, requiring $\sum_{x',y'}M_{x
\otimes y \rightarrow x' \otimes y'}=1$ from probability conservation.
As a result, the second summation in Eq.~(\ref{eq:WAsymptotic3}) is equal to:
\begin{align*}
- \sum_{x,y} E_{x\otimes y} & \lim_{N \rightarrow \infty}
  \Pr(X_{N+1}=x,Y_{N+1}=y) \\
  & = -\sum_{x,y} E_{x\otimes y} \pi_{x\otimes y} \\
  & = -\sum_{x,y} E_{x\otimes y} \pi_{x\otimes y}
  \sum_{x',y'}M_{x \otimes y \rightarrow x' \otimes y'} \\
  & = -\sum_{\substack{x,x', \\ y,y'}} E_{x\otimes y}
  \pi_{x \otimes y} M_{x \otimes y \rightarrow x' \otimes y'}
  ~.
\end{align*}
To compute the first term in Eq.~(\ref{eq:WAsymptotic3}), we do a similar
decomposition. Note that $X_{N+1}$ and $Y'_N$ are determined from $X_N$
and $Y_N$ by iterating with the joint Markov dynamic $M$, and so:
\begin{align}
\label{eq:X'Y'XY}
\Pr(X_{N+1}& = x', Y'_N=y') \nonumber \\
  = & \sum_{x,y}\Pr(X_N=x,Y_N=y)M_{x \otimes y \rightarrow x' \otimes y'}
  ~.
\end{align}
Using Eq.~(\ref{eq:X'Y'XY}) we rewrite the first summation in
Eq.~(\ref{eq:WAsymptotic3}) as:
\begin{align*}
\sum_{x',y'} & E_{x'\otimes y'}
  \lim_{N \rightarrow \infty}\Pr(X_{N+1}=x',Y'_N=y') \\
 = & \sum_{x,y,x',y'} E_{x'\otimes y'} \lim_{N \rightarrow \infty}  \Pr(X_N=x,Y_N=y) M_{x \otimes y \rightarrow x' \otimes y'}\\
 = & \sum_{x,y,x',y'} E_{x'\otimes y'} \pi_{x \otimes y}M_{x \otimes y \rightarrow x' \otimes y'}
   ~. 
\end{align*}
Combining the above, the resulting work production rate is:
\begin{align*}
\langle W \rangle = \sum_{x, x', y, y'}
   (E_{x' \otimes y'} -  E_{x\otimes y})
   \pi_{x \otimes y} M_{x \otimes y \rightarrow x' \otimes y'}
   ~. 
\end{align*}

The same logic leads to the average heat absorption, which turns
out to the same as the work production:
\begin{align*}
\langle Q \rangle =\langle W \rangle
  ~.
\end{align*}
The intuition for this is that these equalities depend on the existence of the
stationary distribution $\pi_{x \otimes y}$ over the ratchet and bit. This is
guaranteed for a finite ratchet with mixing dynamics. Only a finite amount of
energy can be stored in a finite ratchet, so the heat energy flowing in must be
the same as the work flowing out, on the average, to conserve energy. This,
however, may break down with infinite-state ratchets---an important and
intriguing case that our sequels address.

\subsection{Input Process Contributions}

\begin{figure}[tbp]
\centering
\includegraphics[width = 0.9 \columnwidth]{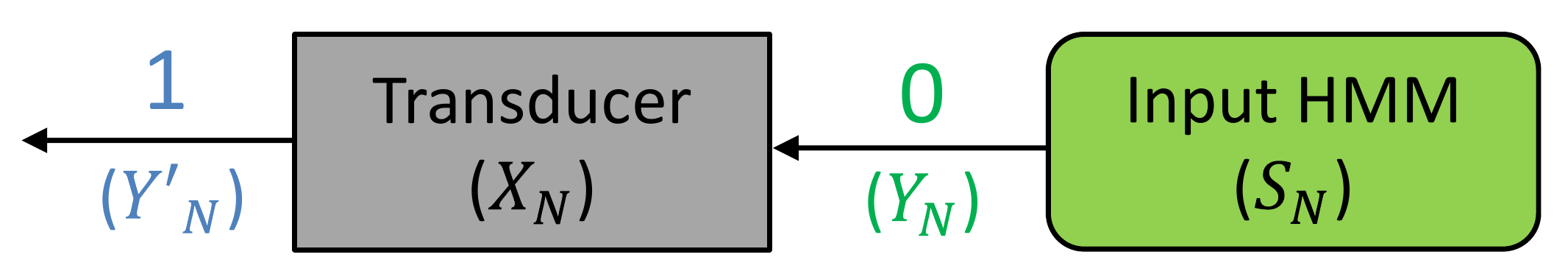}
\caption{\textbf{State variable interdependence:} Input HMM has an autonomous
	dynamics with transitions $S_N \rightarrow S_{N+1}$ leading to input bits
	$Y_N$. That is, $Y_N$ depends only on $S_N$. The joint dynamics of the
	transducer in state $X_N$ and the input bit $Y_N$ leads to the output bit
	$Y'_N$. In other words, $Y'_N$ depend on $X_N$ and $Y_N$ or, equivalently,
	on $X_N$ and $S_N$. Knowing the joint stationary distribution of $X_N$ and
	$S_N$, then determines the stationary distribution of $Y'_N$. However, if
	$Y_N$ and $X_N$ are known, $Y'_N$ is independent of $S_N$.
	}
\label{fig:CompMech_II}
\end{figure}

The results above are expressed in terms of the ratchet, except for the
stationary joint distribution over the input variable and ratchet state:
\begin{align*}
 \pi_{x \otimes y}=\lim_{N \rightarrow \infty} \Pr(X_N=x,Y_N=y)
 ~.
\end{align*}
This quantity is dependent on the input process, as we now describe. We
describe the process generating the input string by an HMM with transition
probabilities:
\begin{align}
T^{(y_N)}_{s_N\rightarrow s_{N+1}}=\Pr(Y_N=y_N,S_{N+1}=s_{N+1}|S_N=s_N)
  ~,
\label{eq:emachine}
\end{align}
where $s_i\in \mathcal{S}$ are the input process' hidden states
~\cite{Boyd15a}. Given that the input HMM is in internal state $s_N$, $T_{s_N
\rightarrow s_{N+1}}^{(y_N)}$ gives the probability to make a transition to the
internal; state $s_{N+1}$ and produce the symbol $y_N$. The dependence between
$X_N$ and $Y_N$ is determined by hidden state $S_N$. So, we rewrite: 
\begin{align*}
& \Pr(X_N = x,Y_N=y) 
   = \sum_s\Pr(X_N=x,Y_N=y,S_N=s) \\ 
  & ~ = \sum_{s}\Pr(Y_N=y|S_N=s)\Pr(X_N=x,S_N=s) \\
  & ~ = \sum_{s, s'}\Pr(Y_N \! = \! y,S_{N+1} \! = \! s'|S_N \! = \! s)
  	\Pr(X_N \! = \! x,S_N \! = \! s) \\ 
  & ~ = \sum_{s, s'} T^{(y)}_{s \rightarrow s'} \Pr(X_N=x,S_N=s)
  ~,
\end{align*}
The second line used the fact that $Y_N$ depends on only $S_N$, as illustrated
in Fig.~\ref{fig:CompMech_II}. The last line used Eq.~(\ref{eq:emachine}).

Combining the above equations gives:
\begin{align*}
\pi_{x \otimes y}
  & = \lim_{N \rightarrow \infty} \Pr(X_N=x,Y_N=y) \\
  & = \lim_{N \rightarrow \infty} \sum_{s,s'}
  T^{(y)}_{s \rightarrow s'} \Pr(X_N=x, S_N=s) \\
  & = \sum_{s,s'}T^{(y)}_{s \rightarrow s'} \pi'_{x \otimes s} ~,~\text{and}\\
  \pi'_{x \otimes s} & =\lim_{N \rightarrow \infty} \Pr(X_N=x, S_N=s)
  ~. 
\end{align*}
Thus, evaluating $\pi_{x \otimes y}$ requires knowing the input process
$T^{(y)}_{s \rightarrow s'}$, which is given, and the stationary joint
distribution $\pi_{x \otimes s}$ over the hidden states and the ratchet states.

To calculate $\pi_{x \otimes s}$, we must consider how $X_{N+1}$ and $S_{N+1}$
are generated from past variables. We notice that the output process is
specified by an HMM whose hidden variables are composed of the hidden variable
of the input HMM and the states of the transducer. In other words, the output
HMM's hidden states belong to the product space $\mathcal{X} \otimes
\mathcal{S}$. As a result, the transition probability of the output HMM is: 
\begin{widetext} 
\begin{align*}
T'^{(y')}_{x \otimes s \rightarrow x'\otimes s'}
  & = \Pr(Y'_N=y',X_{N+1}=x',S_{N+1}=s'|X_N=x,S_N=s) \\
  & = \sum_y \Pr(Y'_N=y',X_{N+1}=x', Y_N = y, S_{N+1}=s'|X_N=x,S_N=s)  \\
  & = \sum_{y}\Pr(Y'_N=y', X_{N+1} = x | Y_N = y, S_{N+1} = s', X_N = x, S_N = s) \Pr(Y_N = y, S_{N+1} = s' | X_N = x, S_N = s) \\ 
  & = \sum_{y}\Pr(Y'_N=y', X_{N+1} = x | Y_N = y, X_N = x) \Pr(Y_N = y, S_{N+1} = s' | S_N = s) \\
  & = \sum_y M_{x \otimes y \rightarrow x' \otimes y'} T^{(y)}_{s \rightarrow s'} \\
  & = \sum_y M_{x  \rightarrow x'}^{(y'|y)} T^{(y)}_{s \rightarrow s'}
  ~,
\end{align*}
\end{widetext}
where the fourth used the facts that $Y'_N$ and $X_{N+1}$ are independent of
$S_N$ and $S_{N+1}$, if $Y_N$ and $X_N$ are known, and $Y_N$ and $S_{N+1}$ are
independent of $X_N$, if $S_N$ is known~\cite{Barn13a,Boyd15a}. Thus, summing
over the output variable $Y'$ yields a Markov dynamic over $\mathcal{X} \otimes
\mathcal{S}$:
\begin{align}
T'_{x \otimes s \rightarrow x' \otimes s'} &= \sum_{y'} T'^{(y')}_{x \otimes s \rightarrow x'\otimes s'}
\\&=\sum_{y,y'} M_{x \otimes y \rightarrow x' \otimes y'} T^{(y)}_{s \rightarrow s'} \nonumber
  ~.
\end{align}
The stationary distribution $\pi'_{x \otimes s}$ is this dynamics' asymptotic
distribution:
\begin{align}
\sum_{x,s}\pi'_{x\otimes s} T'_{x \otimes s \rightarrow x' \otimes s'}=\pi'_{x' \otimes s'}
  ~.
\end{align}
$\pi'$ existence---that is, for a finite state Markov process like $T'$---is
guaranteed by the Perron-Frobenius theorem and it is unique when $T'$ is
ergodic~\cite{VanK92}. In short, we see that $\pi_{x \otimes y}$ is computable
given the ratchet $M_{x \otimes y \rightarrow x' \otimes y'}$ and the input
process generator $T^{(y)}_{s \rightarrow s'}$.

In this way, we derived an expression for the asymptotic work production of an
arbitrary memoryful ratchet with an arbitrary memoryful input process in terms
of HMM generator of the input and the Markovian dynamic over the input bit and
ratchet state. Only a single assumption was made: there is an asymptotic distribution over the the input bit and ratchet state $\pi_{x \otimes y}$.
In summary, there are three steps to calculate the average work production:
\begin{enumerate}
\item Calculate the stationary distribution $\pi'_{x \otimes s}$ over the
	hidden states of the output process $T'^{(y')}_{x \otimes s \rightarrow x'
	\otimes s'}$. The latter which is calculated from the operation of $M_{x
	\otimes y \rightarrow x' \otimes y'}$ on $T^{(y)}_{s \rightarrow s'}$;
\item Use $\pi'$ and $T^{(y)}_{s\rightarrow s'}$ to calculate the stationary
	distribution over the ratchet and input bit at the beginning of the
	interaction interval $\pi_{x \otimes y}$;
\item Using this and the transducer's Markov dynamic, calculate the work
	production:
\begin{align}
\label{eq:WorkProduction}
\langle W \rangle =k_B T \sum_{\substack{x, x', \\ y, y'}}
  \pi_{x \otimes y}
  M_{x \otimes y \rightarrow x' \otimes y'}
  \ln \frac{M_{x' \otimes y' \rightarrow x \otimes y}}{M_{x \otimes y \rightarrow x' \otimes y'}}
  . 
\end{align}
\end{enumerate}

The following Appendix shows how to use this method to calculate average work
production for the specific cases of the period-$2$ environment with and
without phase-slips.

\section{Ratchet Energetics: Specific Expressions}
\label{app:SpecificRatchetEnergetics}

The symbol-labeled transition matrices for the noisy period-$2$ input process
are given by:
\begin{align*}
T^{(0)}&=
\begin{pmatrix}
0 & 0 & 0\\ 
.5 & c & 1-c \\ 
0 & 0 & 0
\end{pmatrix}
\begin{matrix}
D\\ 
E\\ 
F
\end{matrix}
\\ T^{(1)}&=\begin{pmatrix}
0 & 0 & 0\\ 
0 & 0 & 0 \\ 
.5 & 1-c & c
\end{pmatrix}
  ~.
\end{align*}
The transducer form of the ratchet $M$ shown in Fig. 4 is given by the four
conditional symbol-labeled transition matrices:
\begin{align*}
M^{(0|0)}&=
\begin{pmatrix}
0 & \frac{1-\delta}{e} & \gamma\\ 
1-\delta & 0 & 0 \\ 
\delta & 0 & 1-\gamma
\end{pmatrix}
\begin{matrix}
A\\ 
B\\ 
C
\end{matrix}
\\ M^{(1|0)}&=\begin{pmatrix}
0 & 1-\frac{1-\delta}{e} & 0\\ 
0 & 0 & 0 \\ 
0 & 0 & 0
\end{pmatrix}
\\
M^{(0|1)}&=
\begin{pmatrix}
0 & 0 & 0\\ 
1-\frac{1-\delta}{e} & 0 & 0 \\ 
0 & 0 & 0
\end{pmatrix}
\\ M^{(1|1)}&=\begin{pmatrix}
0 & 1-\delta & 0\\ 
\frac{1-\delta}{e} & 0 & \gamma \\ 
0 & \delta & 1-\gamma
\end{pmatrix}
  ~,
\end{align*}
where we switched to the transducer representation of the joint Markov process $M_{x \otimes y \rightarrow x' \otimes y'} = M^{(y'|y)}_{x \rightarrow x'}$ ~\cite{Barn13a,Boyd15a}.

To find the stationary distribution over the causal states of the input bit and
the internal states of the ratchet (step 1), we calculate the output process
$T'^{(y')}_{x \otimes s \rightarrow x' \otimes s'}=\sum_{y}M^{(y'|y)}_{x
\rightarrow x'}T^{(y)}_{s \rightarrow s'}$ and sum over output symbols to get
the Markov dynamic over the hidden states:
\begin{widetext}
\begin{align*}
T' & = T'^{(0)}+T'^{(1)}  = \left(
\begin{array}{ccccccccc}
 0 & 0 & 0 & 0 & 0 & 0 & 0 & 0 & 0 \\
 0 & 0 & 0 & 0.5 & c & \bar{c} & 0.5 \gamma & c \gamma & \gamma\bar{c} \\
 0 & 0 & 0 & 0.5\bar{\delta} & \bar{c} \bar{\delta} & c\bar{\delta} & 0 & 0 & 0 \\
 0 & 0 & 0 & 0 & 0 & 0 & 0 & 0 & 0 \\
 0.5\bar{\delta} & c\bar{\delta} & \bar{c} \bar{\delta} & 0 & 0 & 0 & 0 & 0 & 0 \\
 0.5 & \bar{c} & c & 0 & 0 & 0 & 0.5 \gamma & \gamma\bar{c} & c \gamma \\
 0 & 0 & 0 & 0 & 0 & 0 & 0 & 0 & 0 \\
 0.5 \delta & c \delta & \delta\bar{c} & 0 & 0 & 0 & 0.5\bar{\gamma} & c\bar{\gamma} & \bar{c} \bar{\gamma} \\
 0 & 0 & 0 & 0.5 \delta & \delta\bar{c} & c \delta & 0.5\bar{\gamma} & \bar{c} \bar{\gamma} & c\bar{\gamma} \\
\end{array}
\right)
\begin{matrix}
A\otimes D\\
A \otimes E\\
A \otimes F\\
B \otimes D\\
B \otimes E\\
B \otimes F\\
C \otimes D\\
C \otimes E\\
C \otimes F
\end{matrix}
  ~,
\end{align*}
where $\bar{c} = 1-c $, $\bar{\delta} = 1-\delta $, and $\bar{\gamma} =1- \gamma$.
\end{widetext}

Then, we find the stationary state $\pi'$ over the joint hidden states (step 2), which solves $T' \pi'=\pi'$:
\begin{align*}
\pi'=
\begin{pmatrix}
\pi'_{A \otimes D}\\
\pi'_{A \otimes E}\\
\pi'_{A \otimes F}\\
\pi'_{B \otimes D}\\
\pi'_{B \otimes E}\\
\pi'_{B \otimes F}\\
\pi'_{C \otimes D}\\
\pi'_{C \otimes E}\\
\pi'_{C \otimes F}
\end{pmatrix}
=
\begin{pmatrix}
0\\
\gamma(\delta +c -\delta c)/\nu\\
\gamma(c -\delta c) / \nu \\
0\\
\gamma(c -\delta c) / \nu \\
\gamma(\delta +c -\delta c) / \nu \\
0\\
\delta c / \nu \\
\delta c / \nu
\end{pmatrix}
  ~,
\end{align*}
where $\nu = 2(c\delta+ \gamma (\delta +2c-2\delta c))$.

And, we find the stationary distribution over the ratchet state input bit by
plugging in to the equation $\pi_{x \otimes y}= \sum_{s,s'}T^{(y)}_{s
\rightarrow s'} \pi'_{x \otimes s}$. The result is:
\begin{align*}
\pi=
\begin{pmatrix}
\pi_{A \otimes 0}\\
\pi_{A \otimes 1}\\
\pi_{B \otimes 0}\\
\pi_{B \otimes 1}\\
\pi_{C \otimes 0}\\
\pi_{C \otimes 1}
\end{pmatrix}
=
\begin{pmatrix}
\gamma c / \nu \\
\gamma(\delta +c -2 \delta c) / \nu \\
\gamma(\delta +c -2 \delta c) / \nu \\
\gamma c / \nu \\
\delta c / \nu \\
\delta c / \nu 
\end{pmatrix}
  ~.
\end{align*}

Substituting this stationary distribution into the work expression (step 3) in
Eq. (\ref{eq:WorkProduction}), we find an explicit expression for the
ratchet's work production rate:
\begin{align}
\langle W \rangle =k_B T\frac{(1-\delta)(\delta +c -2\delta c-ec)}{ec \delta/\gamma+  e (\delta + 2c -2 \delta c)}
  ~.
\label{eq:Period2Work}
\end{align}

\subsection{Period-$2$ Input}

To restrict to period-$2$ input sequences with no phase slips we set $c=0$.
Then, $T'$ has the stationary distribution:
\begin{align*}
\pi'_{A\otimes E}=\pi'_{B \otimes F} = 0.5
  ~,
\end{align*}
and all other elements vanish. The ratchet is fully synchronized to the
internal states of the input process. Substituting $c=0$ into Eq.
(\ref{eq:Period2Work}) gives the work production rate when synchronized:
\begin{align*}
\langle W \rangle =k_B T \frac{1-\delta}{e}
  ~.
\end{align*}

\subsection{Noisy Period-$2$ Input}

What happens when the environment fluctuates, generating input sequence phase
slips with probability $c$? Consider the optimal parameter settings at which
the ratchet generates work. When the ratchet behaves as an engine, the optimal
setting is $\gamma = 1$, which follows from the partial derivative of the work
production:
\begin{align*}
\frac{\partial \langle W \rangle }{\partial \gamma }
  = \langle W \rangle
  \frac{e c \delta}{\gamma^2(ec \delta /\gamma+e(\delta +2 c (1-\delta))}
  ~,
\end{align*}
which is always positive when the engine produces work. This means that it is
always possible to enhance our engine's power by increasing $\gamma$ to its
maximum value at $\gamma = 1$. And so, to build an optimal engine that
leverages the noisy period-2 input process, we set $\gamma=1$, yielding:
\begin{align}
\langle W \rangle (\delta, c,\gamma=1)
  = k_B T\frac{(1-\delta)[\delta+c-c(2\delta +e)]}{2ec + \delta e (1-c)}
  ~.
\end{align}

\subsection{Period-2 Input Entropy Rates}

To check that the period-$2$ input process obeys Eq. (\ref{eq:SecondLaw2}), we
calculate the entropy rate:
\begin{align*}
\Delta h_\mu=h'_\mu-h_\mu
  ~.
\end{align*} 
The entropy rate $\hmu$ of a period-$2$ process is:
\begin{align*}
\hmu & = \lim_{N \rightarrow \infty} \frac{\H[Y_{0:N}]}{N} \\
     & =\lim_{N \rightarrow \infty} \frac{1}{N} \nonumber \\
     & = 0 \nonumber
	  ~.
\end{align*}
The entropy rate $\hmu'$ of the output process generated by $T'$ can be calculated
using the uncertainty in the next symbol given the hidden state since
$T'$ is unifilar \cite{Crut01a}:
\begin{align*}
\hmu' & = \lim_{N \rightarrow \infty} \H[Y'_N|S'_N] \\
      & = \lim_{N \rightarrow \infty} \sum_{s'} \H[Y'_N|S'_N=s'] \Pr(S'_N=s')
  ~.
\end{align*}
(No such general expressions hold for nonunifilar transducers.)

For the period-$2$ process, $c=0$, and we see that the stationary state
consists of two states with nonzero probability: $\pi'_{A\otimes E}=\pi'_{B
\otimes F} = 0.5$. These states transition back and forth between each other
periodically, so the current hidden state and output uniquely determine the
next hidden state, meaning this representation is unifilar. Thus, we can use
our calculated output HMM for the entropy rate $\hmu'$.

$A \otimes E$ has probability $\frac{1-\delta}{e}$ of
generating a $1$ and $B \otimes F$ has probability $\frac{1-\delta}{e}$ of
generating a $0$. Thus, the uncertainty in emitting the next bit from either
causal state is:
\begin{align*}
\H[Y'_N|S'_N = A \otimes E]
  & = \H[Y'_N|S'_N = B \otimes F] \\
  & = \H \left( \frac{1-\delta}{e} \right)
  ~.
\end{align*}
Thus, their entropy rates are the same and we find:
\begin{align}
\Delta \hmu = \H \left( \frac{1-\delta}{e} \right)
  ~.
\end{align}

\bibliography{chaos}

\end{document}

%% file: cmechabbrev.tex
\newcommand{\eM}     {\mbox{$\epsilon$-machine}}

\newcommand{\CausalState}   { \mathcal{S} }

\newcommand{\Cmu}       {C_\mu}
\newcommand{\hmu}       {h_\mu}

\newcommand{\forward}{+}
\newcommand{\reverse}{-}
\newcommand{\forwardreverse}{\pm} 

\newcommand{\FutureCausalState} { {\CausalState}^{\forward} }

\newcommand{\PastCausalState}   { {\CausalState}^{\reverse} }

\newcommand{\lastindex}[2]{
  \edef\tempa{0}
  \edef\tempb{#2}
  \ifx\tempa\tempb
    \edef\tempc{#1}
  \else
    \edef\tempa{0}
    \edef\tempb{#1}
    \ifx\tempa\tempb
      \edef\tempc{#2}
    \else
      \edef\tempc{#1+#2}
    \fi
  \fi
  \tempc
}

\newcommand{\CSjoint}[1][,]{
   \edef\tempa{:}
   \edef\tempb{#1}
   \ifx\tempa\tempb
      \ensuremath{\FutureCausalState\!#1\PastCausalState}
   \else
      \ensuremath{\FutureCausalState#1\PastCausalState}
   \fi
}

\newif\ifpm
\edef\tempa{\forwardreverse}
\edef\tempb{\pm}
\ifx\tempa\tempb
   \pmfalse
\else
   \pmtrue
\fi